\documentclass[10pt,british]{article}
\usepackage[utf8]{inputenc}
\usepackage[letterpaper]{geometry}
\usepackage{xcolor}
\usepackage{pdfcolmk}
\usepackage{babel}
\usepackage{float}
\usepackage{amsmath}
\usepackage{amssymb}
\usepackage{graphicx}
\usepackage{esint}
\PassOptionsToPackage{normalem}{ulem}
\usepackage{ulem}
\usepackage[unicode=true,
 bookmarks=false,
 breaklinks=false,pdfborder={0 0 1},colorlinks=true]
 {hyperref}
\hypersetup{
 final,linkcolor=blue,citecolor=blue,filecolor=magenta,urlcolor=blue}

\makeatletter

\providecolor{lyxadded}{rgb}{0,0,1}
\providecolor{lyxdeleted}{rgb}{1,0,0}

\DeclareRobustCommand{\lyxsout}[1]{\ifx\\#1\else\sout{#1}\fi}

\usepackage{xcolor}
\usepackage{babel}

\usepackage[]{ulem}

\usepackage{tabularx}
\usepackage{caption}
\usepackage{sidecap}
\usepackage{enumitem}
\usepackage{notoccite}
\usepackage{cite}
\usepackage{authblk}

\makeatother

\begin{document}
\title{Stochastic entropy production for continuous measurements of an open
quantum system}
\author[1]{D. Matos}
\author[1]{L. Kantorovich}
\author[2]{I.J. Ford}
\affil[1]{\small Department of Physics, King's College London,
	Strand London, WC2R 2LS, United Kingdom}
\affil[2]{\small Department of Physics and Astronomy, University
	College London, Gower Street, London, WC1E 6BT, United Kingdom}

\date{}

\maketitle

\begin{abstract}
We investigate the total stochastic entropy production of a two-level
bosonic open quantum system under protocols of time dependent coupling
to a harmonic environment. These processes are intended to represent
the measurement of a system observable, and consequent selection of
an eigenstate, whilst the system is also subjected to thermalising
environmental noise. The entropy production depends on the evolution
of the system variables and their probability density function, and
is expressed through system and environmental contributions. The continuous
stochastic dynamics of the open system is based on the Markovian approximation
to the exact, noise-averaged stochastic Liouville-von Neumann equation,
unravelled through the addition of stochastic environmental disturbance
mimicking a measuring device. Under the thermalising influence of
time independent coupling to the environment, the mean rate of entropy
production vanishes asymptotically, indicating equilibrium. In contrast,
a positive mean production of entropy as the system responds to time
dependent coupling characterises the irreversibility of quantum measurement,
and a comparison of its production for two coupling protocols, representing
connection to and disconnection from the external measuring device,
satisfies a detailed fluctuation theorem.
\end{abstract}

\section{Introduction}

Most applied and theoretical quantum mechanics research is underpinned
by the theory of open quantum systems. Any realistic quantum system
interacts with its environment, though the interaction in many cases
is weak. Open quantum systems are of interest due to the characteristic
dynamical properties they display, specifically irreversible dissipative
behaviour on approach to a steady state, and decoherence, both of
which are not found in closed systems. These phenomena are of crucial
importance in quantum technological applications such as quantum computing
\cite{shor1995scheme} and in theoretical developments in quantum
thermodynamics \cite{weiss2012quantum}.

In this paper, we concern ourselves with the thermodynamic behaviour
of an open quantum system, more specifically with the stochastic entropy
production associated with the evolution of its reduced density matrix
brought about by changes in interactions with the environment. Such
a framework can naturally describe the consequences arising from a
time dependent Hamiltonian coupling of the system to its environment,
but it can also represent the effect of quantum measurements involving
parts of the environment, such as a measuring device. The field of
stochastic thermodynamics began as a generalisation of the laws of
thermodynamics applied to stochastic systems \cite{sekimoto1998langevin},
such as the behaviour of colloidal particles or molecular systems
exposed to heat baths \cite{seifert2005entropy}. These developments
allow us to compute the entropy production associated with individual
stochastic trajectories of the evolving reduced density matrix. This
stochastic entropy production can satisfy a detailed fluctuation theorem
describing the relationship between the effects of time-reversed versions
of the coupling protocol \cite{spinney2012entropy}.

In order to apply the tools of stochastic thermodynamics to the quantum
regime, a notion of quantum trajectories must be established. Quantum
trajectories have long been used in the field of quantum optics \cite{carmichael2009open},
and they are typically realised through the unravelling of a deterministic
equation of motion for the variables describing the system. Unravelling
is the elaboration of the deterministic dynamics into a stochastic
equation of motion that is capable of describing randomness in system
behaviour, such as fluctuations around the average evolution. The
deterministic dynamics implicitly describes an average over the range
of random behaviour. Randomness might be ascribed to outcomes of continuous
measurements involving the environment \cite{breuer2002theory}, such
as photon counting or homodyne/heterodyne detection, though unitary
evolution of a system together with its environment where the state
of the environment is not fully specified can also be represented
using a framework of stochastic unravelling.

To obtain these unravellings, we start with the stochastic Liouville-von
Neumann (SLN) equation for the system's reduced density matrix \cite{stockburger2001non,stockburger2002exact,stockburger2004simulating,mccaul2017partition,lane2020exactly},
a non-Markovian stochastic differential equation derived from the
Feynman-Vernon path integral-based consideration of open quantum system
behaviour \cite{feynman1963theory}. It is important to note that
the stochasticity in the SLN is merely a mathematical feature that
requires averaging in order to produce the exact deterministic behaviour
of the reduced density matrix. We then derive the Markovian limit
of the noise-averaged SLN equation in order to make possible a simple
unravelling of the deterministic dynamics using a Kraus operator representation
of physical stochasticity brought about by the environment. The stochastic
differential equation that emerges then describes physical randomness
and can provide a basis for deriving the stochastic entropy production.

\begin{figure}[H]
\begin{centering}
\includegraphics[width=0.45\linewidth]{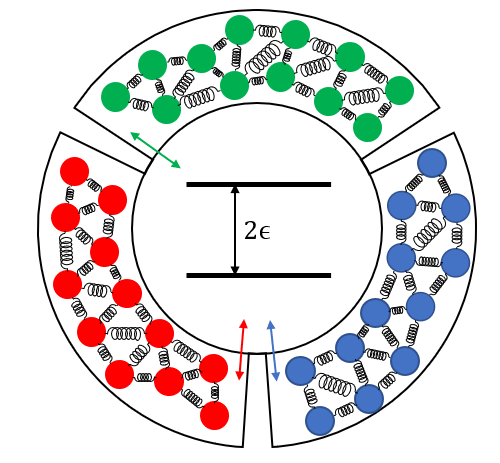}
\par\end{centering}
\caption{An open system consisting of a bosonic two-level system, interacting
with an environment represented by three sets of harmonic oscillators,
each coupled via a different Pauli matrix.\label{fig:An-open-system}}
\label{fig:sys_env_setup}
\end{figure}

We employ stochastic entropy production to quantify the irreversibility
of the continuous measurement of the quantum state of a two-level
bosonic system. We develop a framework where coupling to an external
device causes the system to select an eigenstate of the measured observable,
and subsequent decoupling returns the system to the initial thermal
state. We find that this requires the system to be coupled, initially
with equal strength, to three harmonic baths representing elements
of the environment. Each bath couples to an observable of the system
represented by one of the Pauli matrices. The system and baths are
illustrated in Figure \ref{fig:An-open-system}. The resulting random
exploration of the Bloch sphere, with bias brought about by the system
Hamiltonian, produces behaviour consistent with a thermal Gibbs state
at high temperature. This situation is then disturbed by increasing
one of the environmental coupling strengths, as a representation of
additional interaction associated with a measuring device, which obliges
the system to move towards and dwell in the vicinity of an appropriate
eigenstate according to the Born rule. In our case, we will be dealing
with continuous measurements of the system energy. Returning the coupling
to its initial strength reverses this dynamical behaviour, such that
we can regard the whole sequence as a simple representation of the
dynamics of quantum measurement. Having established the stochastic
dynamics, we can then derive the stochastic entropy production of
measurement using established analysis \cite{spinney2012entropy}
and show that measurement is associated, on average, with positive
entropy production.

Note that incorporating stochasticity into the dynamics is crucial
to computing the irreversibility of a process. It can be demonstrated
that the \emph{mean} evolution of the reduced density matrix towards
stationarity entirely misses the quantum selection of an eigenstate
of the measured observable. Quantifying the irreversibility of a stochastic
trajectory, or indeed the average irreversibility over an ensemble
of trajectories, requires the incorporation of physical randomness
into the dynamics.

The plan for the paper is as follows: in Section \ref{sec:Theory}
the essential ideas of stochastic thermodynamics are introduced, as
well as the SLN equation describing a system coupled to several independent
baths. In Section \ref{sec:System_Setup} we obtain a Markovian approximation
of the noise-averaged SLN equation by taking a high temperature limit
and assuming the relaxation times of the environment to be short.
We then construct a stochastic unravelling of the equation of motion
by choosing appropriate Kraus operators and approximations. In Section
\ref{sec:Results} we study process protocols of variation in the
strength of one of the bath couplings; and then our calculations of
the stochastic entropy production are presented, indicating that there
is an entropic cost of measurement, and showing adherence of the entropy
production to the detailed fluctuation theorem.

\section{Theory\label{sec:Theory}}

Continuous or weak quantum measurements are central to the work we
will be discussing. They bring about a continuous change to the system
being monitored, induced by the environment to which the system is
coupled. Continuous quantum measurements produce evolution described
by a stochastic differential equation (SDE), with an entropy production
which can then be assessed by applying methods of stochastic thermodynamics.
We describe methods for stochastic entropy production in Section \ref{sec:Stochastic_Thermodynamics},
followed by the SLN equation in Section \ref{sec:sln_equation}, which
is the starting point for the dynamics we will be exploring.

\subsection{Stochastic Thermodynamics\label{sec:Stochastic_Thermodynamics}}

The concept of entropy arose over a century ago from a consideration
of the irreversibility of macroscopic phenomena, but our understanding
has evolved significantly since then. Insights into entropy production
were expanded by applying thermodynamic concepts to small and individual
systems in the form of fluctuation theorems. These were first introduced
by Evans \emph{et al.} \cite{evans1993probability,evans1995steady,evans2002fluctuation,carberry2004fluctuations},
and then similar ideas appeared in chaos theory \cite{gallavotti1995dynamical}
and stochastic modelling \cite{kurchan1998fluctuation,lebowitz1999gallavotti},
and were also developed by Jarzynski \cite{jarzynski1997nonequilibrium}
and Crooks \cite{crooks1998nonequilibrium,crooks1999entropy}, amongst
others. These developments introduce the idea of entropy production
associated with individual stochastic trajectories.

Stochastic entropy production can be defined as the contrast in likelihoods
of forward and reverse sequences of system behaviour \cite{seifert2005entropy}.
Specifically, it takes the form of the logarithm of a ratio of the
probabilities of a forward trajectory driven by a \emph{forward} protocol
of driving forces, and the corresponding reversed trajectory driven
by a \emph{reverse} protocol. These protocols are defined such that
one is the time-reversed version of the other, e.g., if the system
Hamiltonian varies in a specific way for the forward protocol, the
exact opposite time dependence takes place in the reverse protocol.
It may be shown that irreversible behaviour such as relaxation towards
a stationary state is then accompanied by positive mean stochastic
entropy production.

We can write the total stochastic entropy production associated with
the evolution of a set of time-dependent coordinates $\boldsymbol{x}(t)$
for $0\le t\le\tau$ as

\begin{equation}
\Delta s_{tot}(\boldsymbol{x})=\ln\frac{p^{F}(\boldsymbol{x}(0),0)\mathcal{P}^{F}(\boldsymbol{x}(\tau)|\boldsymbol{x}(0))}{p^{F}(\boldsymbol{x}(\tau),\tau)\mathcal{P}^{R}(\boldsymbol{x}^{\dagger}(\tau)|\boldsymbol{x}^{\dagger}(0))},\label{eq:entropy_production}
\end{equation}
where $\boldsymbol{x}^{\dagger}$ is a set of coordinates evolving
to form a time-reversed trajectory during the reverse protocol. Concretely,
the reverse trajectory is defined as $\boldsymbol{x}^{\dagger}(t)=\boldsymbol{x}(\tau-t)$,
such that the starting and finishing points $\boldsymbol{x}^{\dagger}(0)$
and $\boldsymbol{x}^{\dagger}(\tau)$ are $\boldsymbol{x}(\tau)$
and $\boldsymbol{x}(0)$ respectively. It is a detailed reversal of
the sequence of events described by the forward trajectory. For simplicity,
our notation disregards an inversion in coordinates that are odd under
time reversal symmetry when defining the reverse trajectory \cite{spinney2012entropy}.
$p^{F}(\boldsymbol{x},t)$ is the probability density function (pdf)
over the coordinates obtained from solving the appropriate Fokker-Planck
equation for the forward protocol, while $\mathcal{P}^{F}$ and $\mathcal{P}^{R}$
are the conditional probability densities for a trajectory from $\boldsymbol{x}(0)$
to $\boldsymbol{x}(\tau)$ under the forward protocol, and for the
time reversed trajectory from $\boldsymbol{x^{\dagger}}(0)$ to $\boldsymbol{x}^{\dagger}(\tau)$
under the reverse protocol, respectively.

Denoting the pdfs of the total entropy production $\Delta s_{tot}$
in forward and reverse driving protocols as $P^{F}\left(\Delta s_{tot}\right)$
and $P^{R}\left(\Delta s_{tot}\right)$, respectively, it is possible
in certain circumstances to relate them as follows \cite{harris2007fluctuation,ford2015stochastic}:

\begin{equation}
e^{\Delta s_{tot}}=\frac{P^{F}\left(\Delta s_{tot}\right)}{P^{R}\left(-\Delta s_{tot}\right)},\label{eq:fluctuation_relation}
\end{equation}
which is called the detailed fluctuation theorem. This result states
that the probability of a negative stochastic entropy production for
a reverse protocol is not zero, in spite of the usual demands of the
second law, though it is exponentially smaller than the probability
of a positive production of the same magnitude during the corresponding
forward protocol.

\subsubsection{SDEs and Fokker-Planck equation\label{sec:intro_fokker_planck_equation}}

We need to construct an appropriate evolution equation for the reduced
density matrix \cite{breuer2002theory}. Kraus operators define mappings
of the density matrix between an initial and final state, and continuous
stochastic trajectories of the density matrix can be generated using
a sequence of Kraus operator maps defined for an infinitesimal time
interval $dt$.

The general Kraus representation of the evolution of a density matrix
$\bar{\rho}$ in a time interval $dt$ is given by

\begin{equation}
\bar{\rho}(t+dt)=\sum_{k}M_{k}\bar{\rho}(t)M_{k}^{\dagger},\label{eq:kraus_representation}
\end{equation}
where $M_{k}$ are the Kraus operators, labelled by the index $k$.
The map should be trace preserving, which requires the operator sum
identity $\sum_{k}M_{k}^{\dagger}M_{k}=\mathbb{I}$ to be satisfied
($\mathbb{I}$ is the identity operator). The Kraus operators depend
on $dt$, and for continuous evolution we require the Kraus operators
to differ incrementally from the identity, i.e. $M_{k}\propto\mathbb{I}$,
when $dt\rightarrow0$.

An interpretation of Eq. \eqref{eq:kraus_representation} is that
each Kraus operator can implement a stochastic action on the system
by the environment, namely
\begin{equation}
\rho(t+dt)=\rho(t)+d\rho=\frac{M_{k}\rho(t)M_{k}^{\dagger}}{{\rm Tr}(M_{k}\rho(t)M_{k}^{\dagger})},\label{eq:change_kraus_operators}
\end{equation}
where $\rho$ is a member of an \emph{ensemble} of density matrices
representing the uncertain current state of the system. The trace
of $\rho$ is clearly preserved. The operation takes place with a
conditional probability given by $p_{k}={\rm Tr}(M_{k}\rho(t)M_{k}^{\dagger})$
such that the average over all possible transformations of $\rho(t)$
is given by
\begin{equation}
\sum_{k}p_{k}\frac{M_{k}\rho(t)M_{k}^{\dagger}}{{\rm Tr}(M_{k}\rho(t)M_{k}^{\dagger})}=\sum_{k}M_{k}\rho(t)M_{k}^{\dagger},\label{eq:average transformation}
\end{equation}
and a further averaging over the ensemble of $\rho(t)$ yields Eq.
\eqref{eq:kraus_representation}, with the over-bar therefore denoting
an ensemble average. For this interpretation to be physically acceptable
$\rho$ must remain positive definite under the mapping Eq. \eqref{eq:kraus_representation}.
This approach can then lead to a Lindblad equation for the ensemble
averaged density matrix $\bar{\rho}$ \cite{jacobs2014quantum}:

\begin{eqnarray}
d\bar{\rho} & = & -i\left[H_{sys},\bar{\rho}\right]dt+\sum_{k}\left(L_{k}\bar{\rho}L_{k}^{\dagger}-\frac{1}{2}\left\{ L_{k}^{\dagger}L_{k},\bar{\rho}\right\} \right)dt,\label{eq:intro_lindblad_equation}
\end{eqnarray}
with Lindblad operators $L_{k}$ related to the $M_{k}.$ Such a deterministic
Lindblad equation can be unravelled into a stochastic differential
equation to simulate stochastic interactions such as continuous measurements
\cite{jacobs2006straightforward}. We show how this framework can
be implemented for our two-level bosonic system in Section \ref{sec:System_Setup}.

To calculate stochastic quantum entropy production, forward and reverse
Kraus operators would be needed to construct forward and reverse stochastic
trajectories, respectively. This was first proposed by Crooks \cite{crooks2008quantum}
by considering forward and reverse changes in the density matrix with
respect to the invariant equilibrium state of the system. This approach
has been used to calculate the entropy production corresponding to
quantum jump unravellings \cite{horowitz2013entropy,leggio2013entropy,elouard2017role,elouard2017probing,monsel2018autonomous},
although the method would not be appropriate for systems without an
equilibrium state \cite{dressel2017arrow,manikandan2019time}. Other
recent developments instead construct the reverse Kraus operators
from the time reversal of the forward operators \cite{dressel2017arrow,manikandan2019time,manikandan2019fluctuation}.

However, we need not seek reverse Kraus operators for situations where
the trajectories are continuous and the stochastic evolution is Markovian.
We need only derive a set of Itô SDEs for the forward dynamics (according
to appropriate forward Kraus operators) from which the entropy production
associated with the evolution of the reduced density matrix may be
computed using the approach developed in \cite{spinney2012entropy}.
We could construct SDEs for each element of the density matrix, though
it is more convenient, and physically more transparent, to consider
the dynamics of quantum expectation values of various physical operators,
as we shall see.

Together with the set of SDEs, we require an associated Fokker-Planck
equation for the pdf of the chosen system variables. Let us therefore
consider an Itô SDE for a vector of dynamical variables $\boldsymbol{x}$
involving a vector of Wiener increments $\boldsymbol{dW}$, a vector
of deterministic rate functions $\boldsymbol{A}$ and a matrix of
noise coefficients $\boldsymbol{B}$:

\begin{equation}
d\boldsymbol{x}=\boldsymbol{A}(\boldsymbol{x},t)dt+\boldsymbol{B}(\boldsymbol{x},t)\boldsymbol{dW}.\label{eq:general_SDE}
\end{equation}
The corresponding Fokker-Planck equation that describes the evolution
of the pdf $p(\boldsymbol{x},t)$ is \cite{risken1996fokker}

\begin{equation}
\frac{\partial p(\boldsymbol{x},t)}{\partial t}=\sum_{i}\frac{\partial}{\partial x_{i}}\left[-A_{i}(\boldsymbol{x},t)p(\boldsymbol{x},t)+\sum_{j}\frac{\partial}{\partial x_{j}}\left(D_{ij}(\boldsymbol{x},t)p(\boldsymbol{x},t)\right)\right],\label{eq:fokker_planck_eq_multidim}
\end{equation}
where $\boldsymbol{D=\frac{1}{2}\boldsymbol{B}\boldsymbol{B}^{T}}$
is the diffusion matrix. We shall find that $\boldsymbol{D}$ is diagonal
for our model which simplifies matters.

\subsubsection{Stochastic entropy production\label{sec:stochastic_entropy_production}}

From the definition in Eq. \eqref{eq:entropy_production}, it is possible
to divide the total entropy production into two contributions associated
with the environment and the system, respectively. This division is
somewhat arbitrary and is introduced largely for ease of interpreting
the behaviour. We calculate the total entropy production using the
approach of Ref. \cite{spinney2012entropy}, first separating the
deterministic part of Eq. \eqref{eq:general_SDE} into two contributions:

\begin{equation}
d\boldsymbol{x}=\boldsymbol{A}^{irr}(\boldsymbol{x},t)dt+\boldsymbol{A}^{rev}(\boldsymbol{x},t)dt+\boldsymbol{B}(\boldsymbol{x},t)\boldsymbol{dW},
\end{equation}
where $\boldsymbol{A}^{irr}$ and $\boldsymbol{A}^{rev}$ are the
irreversible and reversible contributions to $\boldsymbol{A}$, respectively.
Specifically, if $dx_{i}$ (the $i$-th component of $d\boldsymbol{x}$)
is even with respect to time reversal, then $A_{i}^{rev}dt$ is also
even. Since $dt$ is odd, $A_{i}^{rev}$ has to be odd: $A_{i}^{rev}$
transforms in the opposite way to $x_{i}$ under the time reversal.
By similar reasoning $A_{i}^{irr}$ transforms in the same way as
$x_{i}$. This separation is necessary for the calculation of the
entropy production \cite{spinney2012entropy}.

Explicit expressions for calculating the two contributions of the
entropy production have been derived. For a system with a diagonal
diffusion matrix $\boldsymbol{D}$, the incremental \emph{environmental}
stochastic entropy production is \cite{spinney2012entropy}:

\begin{align}
d\Delta s_{env} & =\sum_{i=1}^{N}\left[\frac{A_{i}^{ir}}{D_{ii}}dx_{i}-\frac{A_{i}^{rev}A_{i}^{irr}}{D_{ii}}dt+\frac{\partial A_{i}^{irr}}{\partial x_{i}}dt-\frac{\partial A_{i}^{rev}}{\partial x_{i}}dt-\frac{1}{D_{ii}}\frac{\partial D_{ii}}{\partial x_{i}}dx_{i}\right.\label{eq:stochastic_environmental_entropy}\\
 & \left.\qquad\qquad\qquad+\frac{A_{i}^{rev}-A_{i}^{irr}}{D_{ii}}\frac{\partial D_{ii}}{\partial x_{i}}dt-\frac{\partial^{2}D_{ii}}{\partial x_{i}^{2}}dt+\frac{1}{D_{ii}}\left(\frac{\partial D_{ii}}{\partial x_{i}}\right)^{2}dt\right],\nonumber 
\end{align}
where $N$ is the number of dynamical variables contributing to the
entropy production. Note that this expression \eqref{eq:stochastic_environmental_entropy}
contains Wiener increments and hence the environmental entropy production
$\Delta s_{env}$ is explicitly a stochastic variable.

The second contribution, the incremental \emph{system} stochastic
entropy production, is related to the change in the logarithm of the
pdf $p(\boldsymbol{x},t)$ of the system, obtained from the Fokker-Planck
equation:

\begin{equation}
\begin{aligned}d\Delta s_{sys} & =-d\left(\ln p(\boldsymbol{x},t)\right)\\
 & =-\frac{\partial\ln p(\boldsymbol{x},t)}{\partial t}dt-\sum_{i}\frac{\partial\ln p(\boldsymbol{x},t)}{\partial x_{i}}dx_{i}-\sum_{i}D_{ii}\frac{\partial^{2}\ln p(\boldsymbol{x},t)}{\partial x_{i}^{2}}dt,
\end{aligned}
\label{eq:stochastic_system_entropy}
\end{equation}
where the second line is obtained using Itô's Lemma. The total stochastic
entropy production is then given by the sum of the environmental and
system contributions, $d\Delta s_{tot}=d\Delta s_{sys}+d\Delta s_{env}$.
From these individual stochastic contributions, it is possible to
obtain the averaged system and environmental entropy productions by
cumulative summing up of their respective incremental stochastic contributions,
followed by the averaging over many realisations of the noises and
hence system evolution.

The averaged total entropy production also has an analytical form.
For certain circumstances, namely sufficient elapsed time for the
pdf to become time independent, and the maintenance of a condition
of detailed balance, this average is given by the Kullback--Leibler
divergence between the initial and final pdfs \cite{spinney2012entropy}:

\begin{equation}
\langle\langle\Delta s_{tot}\rangle\rangle=\int d\boldsymbol{x}\,p(\boldsymbol{x},t_{init})\ln\frac{p(\boldsymbol{x},t_{init})}{p(\boldsymbol{x},t_{final})}.\label{eq:averaged_total_entropy_production}
\end{equation}
The double angled brackets denote averages over all noise histories
and initial coordinates, over the time interval from $t_{init}$ to
$t_{final}$ arising from all initial states in the ensemble. Likewise,
the averaged \emph{system} entropy production also has an analytical
form, usually taken to be

\begin{equation}
\langle\langle\Delta s_{sys}\rangle\rangle=\Delta S_{G}=-\int d\boldsymbol{x}\,p(\boldsymbol{x},t_{final})\,\ln p(\boldsymbol{x},t_{final})+\int d\boldsymbol{x}\,p(\boldsymbol{x},t_{init})\,\ln p(\boldsymbol{x},t_{init}),\label{eq:averaged_system_entropy_production}
\end{equation}
namely the difference in the Gibbs entropy of the final and initial
probability distributions. This result again emerges only under certain
conditions, and we investigate additional contributions in Section
\ref{sec:boundary_term_effects}.

\subsection{Stochastic Liouville-von Neumann equation\label{sec:sln_equation}}

Having described some of the background concepts of stochastic entropy
production, we now introduce the stochastic dynamics under consideration.
As briefly discussed in the introduction, a non-Markovian SLN equation
can provide an exact treatment of the dynamics of the open system
reduced density matrix, implicitly after averaging over all manifestations
of the noises associated with the environment (see below). The approach
involves taking an average over stochastic solutions of the equation
resulting in a deterministic trajectory from which physical predictions
can be obtained \cite{lane2020exactly,matos2020efficient}.

However, in order to unravel the SLN equation in a straightforward
way, we need to start from a Markovian limit of the system behaviour,
such that the system interacts with an environment that does not possess
any memory. Another perspective is to consider the environmental correlation
times, which in a Markovian limit are very short when compared to
the characteristic timescales for the dynamics of the open system.
In this section we obtain the deterministic Markovian equation of
motion for the physically meaningful, ensemble averaged reduced density
matrix of our system. In Section \ref{sec:System_Setup} that follows,
the unravelling procedure is considered.

Many non-Markovian methods for open quantum systems begin their development
with the full density matrix of the environment and system and proceed
to take a partial trace in order to derive an equation of motion exclusively
for the reduced density matrix of the system. The Feynman-Vernon influence
functional formalism is one such method, where the reduced density
matrix of the open system is represented as a path integral over all
environmental modes made up of an infinite number of harmonic oscillators
\cite{feynman1963theory}. Methods which rely on this include quasi-adiabatic
path integrals \cite{makri1995tensor}, hierarchical equations of
motion \cite{yan2004hierarchical,suess2014hierarchy}, stochastic
Schrödinger equations \cite{orth2013nonperturbative} and the Stochastic
Liouville-von Neumann (SLN) equation and its variations \cite{stockburger2001non,stockburger2002exact,stockburger2004simulating,mccaul2017partition,lane2020exactly}.
Other methods that do not rely on the Feynman-Vernon functional include
projection operator methods such as the Nakajima-Zwanzig equation
\cite{nakajima1958quantum,zwanzig1960ensemble,mori1965transport}.

The SLN equation is a stochastic differential equation containing
complex cross-correlated coloured noises representing the environment;
to arrive at the appropriate deterministic description, an average
needs to be taken over such noises. Note again that the outcome of
such a process will describe the evolution of an average of an ensemble
of physical density matrices, according to the framework of interpretation
set out in Section \ref{sec:intro_fokker_planck_equation}.

The SLN equation describes an open quantum system, with coordinates
$q$, coupled to a large number of harmonic oscillators that represent
the environment. Here we shall consider the system to be coupled to
three independent sets of oscillators; the reason for doing so will
become apparent. The SLN formalism \cite{stockburger2002exact} can
easily be extended to accommodate this. The Hamiltonian coupling of
the system to the $k^{th}$ set of oscillators is taken to be linear
in the oscillator coordinates $\xi_{ik}$, where $i$ labels the oscillators,
but it can depend on a general coupling operator $f_{k}(q)$ of the
system coordinates $q$. The full Hamiltonian can be written as follows

\begin{equation}
H_{total}(q,\{\xi_{ik}\},t)=H_{sys}(q,t)+\sum_{k}\left[H_{env;k}(\xi_{ik})-\sum_{i}\xi_{ik}f_{k}(q,t)\right],\label{eq:general_hamiltonian}
\end{equation}
where $H_{env;k}(\xi_{ik})$ is the Hamiltonian for the $k^{th}$
set of environmental oscillators, and $H_{sys}$ is the Hamiltonian
of the open system. To derive the SLN equation, the initial density
matrix of the full system is taken to be the tensor product of the
reduced density matrix of the system $\rho_{q}(t_{0})$ and the environmental
density matrix $\rho_{\xi}(t_{0})$, i.e. $\rho_{0}=\rho_{q}(t_{0})\otimes\rho_{\xi}(t_{0})$,
where $t_{0}$ is an initial time. We shall be considering a weak
coupling limit so this partitioned state is an appropriate approximation
for the system we are investigating.

Based on the Hamiltonian of Eq. \eqref{eq:general_hamiltonian}, it
is possible to obtain a stochastic differential equation for the (unphysical)
stochastic reduced density matrix $\rho_{s}(t)$ of the system driven
by a pair of complex coloured Gaussian noises, $\eta_{k}(t)$, and
$\nu_{k}(t)$, for each oscillator set $k$ \cite{stockburger2001non,stockburger2002exact,stockburger2004simulating,mccaul2017partition}.
It should be stressed again that these stochastic noises are purely
mathematical constructs, and do not generate individual physical trajectories
\cite{mccaul2017partition,lane2020exactly}. Physical results only
arise when the noise-driven trajectories are averaged over their many
realisations. The SLN equation (with $\hbar=1$) becomes:

\begin{equation}
\frac{d\rho_{s}(t)}{dt}=-i\left[H_{sys},\rho_{s}(t)\right]+i\sum_{k}\left(\eta_{k}(t)\left[f_{k},\rho_{s}(t)\right]+\frac{\nu_{k}(t)}{2}\left\{ f_{k},\rho_{s}(t)\right\} \right),\label{eq:SLN_equation_several_couplings}
\end{equation}
where the square brackets correspond to a commutator, and the curly
brackets to an anticommutator. Each set of environmental oscillators
is connected independently to the open system and the associated noises
satisfy the following correlations:

\begin{equation}
\langle\eta_{k}(t)\eta_{k}(t^{\prime})\rangle=\int_{0}^{\infty}\frac{d\omega}{\pi}J_{k}(\omega)\coth\left(\frac{1}{2}\beta_{k}\omega\right)\cos\left(\omega\left(t-t^{\prime}\right)\right)\equiv K_{k}^{\eta\eta}(t-t^{\prime}),\label{eq:kernel_etaeta}
\end{equation}
\begin{equation}
\langle\eta_{k}(t)\nu_{k}(t^{\prime})\rangle=-2i\Theta(t-t^{\prime})\int_{0}^{\infty}\frac{d\omega}{\pi}J_{k}(\omega)\sin\left(\omega\left(t-t^{\prime}\right)\right)\equiv K_{k}^{\eta\nu}(t-t^{\prime}),\label{eq:kernel_etanu}
\end{equation}
where $J_{k}(\omega)$ is the spectral density of the $k^{th}$ oscillator
set, $\beta_{k}=1/k_{B}T_{k}$ is its inverse temperature (with $k_{B}$
being set to 1 hereafter), $\Theta(t-t^{\prime})$ the Heaviside function,
and the angled brackets denote an average over the environmental noises;
all other correlation functions are zero. There exist several ways
to construct the coloured noises, with each scheme affecting the convergence
of results in different ways \cite{matos2020efficient}.

\subsection{Noise-averaged SLN equation\label{sec:Averaged_SLN}}

We need to average Eq. \eqref{eq:SLN_equation_several_couplings}
with respect to realisations of the environmental noises (to be denoted
by single angle brackets) to obtain an exact and deterministic evolution
equation for the physical reduced density matrix of the open system.
Taking the average of both sides of Eq. \eqref{eq:SLN_equation_several_couplings}
and denoting the physical density matrix averaged over the noises
as $\bar{\rho}=\langle\rho_{s}\rangle$, we write:

\begin{equation}
\frac{d\bar{\rho}(t)}{dt}=-i\left[H_{sys},\bar{\rho}(t)\right]+i\sum_{k}\left(\left[f_{k},\langle\eta_{k}(t)\rho_{s}(t)\rangle\right]+\frac{1}{2}\left\{ f_{k},\langle\nu_{k}(t)\rho_{s}(t)\rangle\right\} \right).\label{eq:averaged_SLN_equation_several_couplings}
\end{equation}
In order to calculate the averages of the stochastic density matrix
multiplied by the noises that appear in the right hand side, the Furutsu-Novikov
theorem may be used \cite{furutsu1964statistical,novikov1965functionals}.
For a set of noises $\zeta_{i}$ with the correlation

\begin{equation}
\langle\zeta_{i}(t)\zeta_{j}(t^{\prime})\rangle=F_{ij}(t,t^{\prime}),
\end{equation}
the Furutsu-Novikov theorem states that

\begin{equation}
\langle\zeta_{i}(t)A[\zeta]\rangle=\sum_{j}\int_{0}^{t}dt^{\prime}F_{ij}(t,t^{\prime})\left\langle \frac{\delta A[\phi]}{\delta\zeta_{j}(t^{\prime})}\right\rangle ,\label{eq:novikov_theorem}
\end{equation}
where $A[\zeta]$ is a functional of the noises and $\delta A[\zeta]/\delta\zeta_{j}$
is its functional derivative with respect to $\zeta_{j}$. The averaged
product of a noise and a noise-dependent functional may therefore
be transformed into an integral. We use Eqs. \eqref{eq:novikov_theorem}
in \eqref{eq:averaged_SLN_equation_several_couplings} to obtain the
required explicit expressions for the averages of the noises multiplied
by $\rho_{s}$:
\begin{equation}
\begin{aligned}\frac{d\bar{\rho}(t)}{dt} & =-i\left[H(t),\bar{\rho}(t)\right]+i\sum_{k}\left[f_{k},\int_{0}^{t}dt^{\prime}K_{k}^{\eta\eta}(t-t^{\prime})\left\langle \frac{\delta\rho_{s}(t)}{\delta\eta_{k}(t^{\prime})}\right\rangle +\int_{0}^{t}dt^{\prime}K_{k}^{\eta\nu}(t-t^{\prime})\left\langle \frac{\delta\rho_{s}(t)}{\delta\nu_{k}(t^{\prime})}\right\rangle \right]\\
 & \qquad\qquad\qquad\qquad\qquad\qquad+\frac{i}{2}\sum_{k}\left\{ f_{k},\int_{0}^{t}dt^{\prime}K_{k}^{\eta\nu}(t^{\prime}-t)\left\langle \frac{\delta\rho_{s}(t)}{\delta\eta_{k}(t^{\prime})}\right\rangle \right\} .
\end{aligned}
\label{eq:averaged_sln}
\end{equation}
From Eq. \eqref{eq:kernel_etanu} we observe the presence of the Heaviside
function $\Theta(t-t^{\prime})$, which vanishes for $t^{\prime}>t.$
The integrals in Eq. \eqref{eq:averaged_sln} have an upper limit
$t$, which implies that $K_{k}^{\eta\nu}(t^{\prime}-t)=0$ in the
last term since $t^{\prime}$ is always smaller than $t$ due to the
integration limits. Hence, the last term is zero and can be dropped;
only the commutator terms remain. We then need to calculate the functional
derivatives, and after some manipulation, see Appendix \ref{subsec:Averaged-SLN-equation},
we obtain

\begin{align}
\frac{d\bar{\rho}(t)}{dt} & =-i\left[H_{sys},\bar{\rho}(t)\right]-\int_{0}^{t}dt^{\prime}K^{\eta\eta}(t-t^{\prime})\sum_{k}\left[f_{k},\left\langle U_{+}(t,t^{\prime})f_{k}(t^{\prime})U_{+}(t^{\prime},t)\rho_{s}(t)-\rho_{s}(t)U_{-}(t,t^{\prime})f_{k}(t^{\prime})U_{-}(t^{\prime},t)\right\rangle \right]\label{eq:averaged_sln_general}\\
 & \qquad\qquad\qquad+\frac{1}{2}\int_{0}^{t}dt^{\prime}K_{k}^{\eta\nu}(t-t^{\prime})\sum_{k}\left[f_{k},\left\langle U_{+}(t,t^{\prime})f_{k}(t^{\prime})U_{+}(t^{\prime},t)\rho_{s}(t)+\rho_{s}(t)U_{-}(t,t^{\prime})f_{k}(t^{\prime})U_{-}(t^{\prime},t)\right\rangle \right],\nonumber 
\end{align}
where $U_{+}$ and $U_{-}$ are forward and reverse propagators, respectively
(see Appendix \ref{subsec:Averaged-SLN-equation}). Eq. \eqref{eq:averaged_sln_general}
is the most general form of the averaged SLN equation, as no approximations
have been made so far. Note its non-Markovian nature. Also note that
it describes the evolution of a reduced density matrix averaged over
physical fluctuations, along the lines of Eq. \eqref{eq:intro_lindblad_equation}.
However, this equation still contains the stochastic density matrix
$\rho_{s}$ and hence is not a self-contained equation for the physical
ensemble-averaged density matrix $\bar{\rho}$. To be able to work
with an equation depending exclusively on the latter, some simplifications
need to be made; this will be done in the next Section.

\subsection{Markovian limit of the SLN equation\label{sec:Markovian_limit}}

Eq. \eqref{eq:averaged_sln_general} in its exact non-Markovian form
is difficult to work with as it still contains the stochastic density
matrix. A self-contained equation for the physical density matrix
$\bar{\rho}(t)$ can be obtained, however, in the Markovian limit
by approximating certain properties of the environment. We begin by
assuming that all the sets of oscillators coupled to the system have
the same temperature $T_{k}\equiv T$, and the same spectral density.
This is not a necessary assumption or approximation in our analysis,
but it will simplify our notations. Next we take the limit in which
the temperature of the environment $T$ is very large such that $T$
is much greater than any characteristic energy quantum of its harmonic
oscillators, $\beta\omega\ll1$, and we will also adopt the same Ohmic
spectral density $J(\omega)=\alpha\omega$ for all oscillator sets,
where $\alpha$ is a proportionality constant. In this case we can
use the first order expansion of $\coth\left(\frac{\beta\omega}{2}\right)\approx\frac{2}{\beta\omega}$,
following \cite{yan2016stochastic}, and obtain the following results
for the correlation functions of Eqs. \eqref{eq:kernel_etaeta} and
\eqref{eq:kernel_etanu}:

\begin{equation}
K^{\eta\eta}(t-t')\approx\int_{0}^{\infty}\frac{d\omega}{\pi}\alpha\omega\frac{2}{\beta\omega}\cos\left(\omega(t-t^{\prime})\right)=\frac{2\alpha}{\beta}\delta(t-t^{\prime}),\label{eq:markovian_ketaeta}
\end{equation}
and
\begin{equation}
\begin{aligned}K^{\eta\nu}(t-t') & =2i\alpha\Theta(t-t^{\prime})\frac{\partial}{\partial t}\left[\frac{1}{\pi}\int_{0}^{\infty}d\omega\cos\left(\omega(t-t^{\prime})\right)\right]=-2i\alpha\Theta(t-t^{\prime})\frac{\partial}{\partial t^{\prime}}\left[\frac{1}{\pi}\int_{0}^{\infty}d\omega\cos\left(\omega(t-t^{\prime})\right)\right]\\
 & =-2i\alpha\Theta(t-t^{\prime})\frac{\partial}{\partial t^{\prime}}\delta(t-t^{\prime}).
\end{aligned}
\label{eq:markovian_ketanu}
\end{equation}
Note that the correlation functions are the same for all the oscillator
sets, hence we have dropped the index $k$ here. Eqs. \eqref{eq:markovian_ketaeta}
and \eqref{eq:markovian_ketanu} appear in \cite{caldeira1983path},
although a different route to obtain them was taken, specifically
that of introducing a cutoff in the spectral density and allowing
it to be much larger than the dynamical timescales of the system.
Inserting Eqs. \eqref{eq:markovian_ketaeta} and \eqref{eq:markovian_ketanu}
into Eq. \eqref{eq:averaged_sln_general} leads to the Markovian limit
of the averaged SLN equation that contains only $\bar{\rho}(t)$.
The equation takes a similar form to that found in other work \cite{yan2016stochastic}:

\begin{align}
\frac{d\bar{\rho}(t)}{dt} & =-i\left[H_{sys},\bar{\rho}(t)\right]-\sum_{k}\left[f_{k},\frac{\alpha}{\beta}\left[f_{k},\bar{\rho}(t)\right]-\frac{\alpha}{2}\left\{ \left[H_{sys},f_{k}\right],\bar{\rho}(t)\right\} +\frac{i\alpha}{2}\left\{ \frac{\partial f_{k}}{\partial t},\bar{\rho}(t)\right\} \right]\nonumber \\
 & =-i\left[H_{sys},\bar{\rho}(t)\right]+\sum_{k}\left(-\frac{\alpha}{\beta}\left[f_{k},\left[f_{k},\bar{\rho}(t)\right]\right]+\frac{\alpha}{2}\left[f_{k},\left\{ \left[H_{sys},f_{k}\right],\bar{\rho}(t)\right\} \right]-\frac{i\alpha}{2}\left[f_{k},\left\{ \frac{\partial f_{k}}{\partial t},\bar{\rho}(t)\right\} \right]\right),\label{eq:markovian_limit_sln_general}
\end{align}
but unlike the result published in \cite{yan2016stochastic}, this
equation contains a term involving the time dependence of the system
coupling operators $f_{k}$.

\section{System setup and equations of motion\label{sec:System_Setup}}

In the last section we obtained the Markovian limit of the evolution
equation for the ensemble averaged reduced density matrix. Here we
shall unravel this equation, namely add noise terms, such that individual
trajectories correspond to different physical realisations of the
quantum state diffusion of the system under the influence of an underspecified
environment.

\subsection{Setup of the system and the unravelling procedure}

\subsubsection{Two-level system and choice of environments\label{sec:spinboson}}

We consider a two-level bosonic system with Hamiltonian

\begin{equation}
H_{sys}=\epsilon\sigma_{z},\label{eq:system hamiltonian}
\end{equation}
where $\sigma_{z}$ is the usual Pauli matrix and $\epsilon$ has
the dimension of energy. We take the system Hamiltonian to be constant
in time. We require the environmental interaction with the open system
to be consistent with a thermal state subject to our earlier assumptions
of weak coupling and high temperature. This means that the insertion
of $\bar{\rho}\propto\mathbb{I}-\beta H_{sys}$ should cause the right
hand side of Eq. \eqref{eq:markovian_limit_sln_general}, with time
independent $f_{k}$, to vanish:
\begin{align}
\frac{d\bar{\rho}}{dt} & \propto-i\left[H_{sys},\mathbb{I}-\beta H_{sys}\right]+\frac{\alpha}{\beta}\sum_{k}\left(-\left[f_{k},\left[f_{k},\mathbb{I}-\beta H_{sys}\right]\right]+\frac{1}{2}\left[f_{k},\left\{ \left[\beta H_{sys},f_{k}\right],\mathbb{I}-\beta H_{sys}\right\} \right]\right)\nonumber \\
 & =-\frac{\alpha\beta}{2}\sum_{k}\left[f_{k},\left\{ \left[H_{sys},f_{k}\right],H_{sys}\right\} \right].\label{eq:thermal limit}
\end{align}
The anticommutator vanishes exactly for our choice of Hamiltonian
where $H_{sys}^{2}\propto\mathbb{I}$ and the implication is that
any choice of the operators $f_{k}$, through which the system interacts
with the environment, is consistent with the appropriate stationary
thermal state. Since the $f_{k}$ are hermitian according to the structure
of the Hamiltonian in Eq. \eqref{eq:general_hamiltonian}, it is natural,
in our case, to employ three coupling operators proportional to the
Pauli matrices. Furthermore, it is natural for the initial strength
of the coupling to be the same for each operator, in order that the
system-environment interaction should be isotropic. Note that the
same framework, if needed, could then be employed for system Hamiltonians
proportional to $\sigma_{x}$ or $\sigma_{y}$ as well as the choice
in Eq. \eqref{eq:system hamiltonian}.

This coupling scheme has the added advantage that it allows us to
model the act of measurement of the system energy in a straightforward
way, merely by elevating the strength of the coupling to the operator
$\sigma_{z}$. This represents an additional interaction between the
system and an external measuring device, and in the next section we
shall see that the unravelled system dynamics under such an interaction
can indeed correspond to the stochastic selection of one of the energy
eigenstates. Specifically, the Hamiltonian, Eq. \eqref{eq:system hamiltonian},
has two eigenvectors $\vert\pm\epsilon\rangle$ with eigenvalues $\pm\epsilon$,
respectively, and the environmental coupling to $\sigma_{z}$ drives
the density matrix stochastically towards the form $\vert+\epsilon\rangle\langle+\epsilon\vert$
or $\vert-\epsilon\rangle\langle-\epsilon\vert$. However, the scheme
of coupling to the three Pauli matrices means that these eigenstates
are not fixed points of the dynamics: the tendency is only for the
system to dwell in the vicinity of one or other of the eigenstates,
to an extent that depends on the strength of environmental coupling
to $\sigma_{z}$.

To summarise, we shall couple three independent sets of harmonic oscillators
to our open system using operators:

\begin{equation}
f_{x}=\gamma_{0}\sigma_{x}\,,\,\,\,f_{y}=\gamma_{0}\sigma_{y}\,\,\,\text{and}\,\,\,f_{z}=\gamma(t)\sigma_{z},\label{eq:couplings_environments}
\end{equation}
where for simplicity we replace labels $k$ by the Cartesian indices
$x$, $y$ and $z$. $\gamma_{0}$ is the coupling coefficient between
the system and the environment, and $\delta\gamma(t)=\gamma(t)-\gamma_{0}$
is the coupling coefficient between the system and the device measuring
$H_{sys}$. If $\gamma(t)$ is increased from an initial value of
$\gamma_{0}$, we can model measurement of the system energy while
under the influence of a thermal bath. We can calculate the stochastic
entropy production related to this process. Furthermore, we can model
the detachment of the measuring device by a protocol where $\gamma(t)$
returns to $\gamma_{0}$ allowing the system to re-thermalise, and
again compute the associated entropy production.

Note also that, with this choice of the coupling, the last term in
Eq. \eqref{eq:markovian_limit_sln_general} containing the time derivative
of the $f_{k}$ vanishes exactly; this is obvious for constant $f_{x}$
and $f_{y}$, while for $f_{z}$ it follows from the fact that $\partial f_{k}/\partial t$
is proportional to $\sigma_{z}$.

\subsubsection{Unravelled stochastic equation}

To obtain a form of Eq. \eqref{eq:markovian_limit_sln_general} that
allows for a straightforward stochastic unravelling, we would need
to write it in a Lindblad form with positive coefficients \cite{breuer2002theory}.
It would be possible to do so by extending the Hilbert space and then
adding stochasticity \cite{breuer1999stochastic,breuer2004genuine},
though that approach will not be taken here. Instead we construct
Lindblad operators that match the dynamics of Eq. \eqref{eq:markovian_limit_sln_general}
in accordance with our high temperature approximation $\beta H_{sys}\ll1$.

We shall now show that the Markovian averaged (deterministic) SLN
Eq. \eqref{eq:markovian_limit_sln_general} is equivalent, to lowest
order in $\beta\epsilon$, to the Lindblad equation when an appropriate
choice of the Lindblad operators is made. Indeed, consider the Lindblad
equation \eqref{eq:intro_lindblad_equation} using operators

\begin{equation}
L_{k}=\lambda\left(f_{k}-\frac{1}{4}\left[\beta H_{sys},f_{k}\right]\right),\label{eq:lindblad_operator}
\end{equation}
where $\lambda=\sqrt{\frac{2\alpha}{\beta}}$. By inserting these
operators into Eq. \eqref{eq:intro_lindblad_equation} we obtain

\begin{alignat}{1}
d\bar{\rho} & =-i\left[H_{sys}-\frac{i\alpha}{4}\sum_{k}\left[H_{sys},f_{k}^{2}\right],\bar{\rho}\right]dt+\sum_{k}\left(-\frac{\lambda^{2}}{2}\left[f_{k},\left[f_{k},\bar{\rho}\right]\right]+\frac{\lambda^{2}}{4}\beta\left[f_{k},\left\{ \left[H_{sys},f_{k}\right],\bar{\rho}\right\} \right]\right)dt\label{eq:markovian_limit_sln}\\
 & \qquad\qquad\qquad\qquad\qquad-\frac{\lambda^{2}}{16}\sum_{k}\beta^{2}\left(\left[H_{sys},f_{k}\right]\bar{\rho}\left[H_{sys},f_{k}\right]-\frac{1}{2}\left\{ \left[H_{sys},f_{k}\right]^{2},\bar{\rho}\right\} \right)dt.\nonumber 
\end{alignat}
We note that $\left[H_{sys},f_{k}^{2}\right]$ vanishes when $f_{k}^{2}\propto\mathbb{I}$,
which holds in our case. The final contribution on the right hand
side is a factor $\beta H_{sys}$ smaller in magnitude than the penultimate
term, and therefore can be neglected. With these considerations we
recover the Markovian averaged SLN Eq. \eqref{eq:markovian_limit_sln_general}
confirming the suitability of the Lindblad operators in Eq. \eqref{eq:lindblad_operator}.

Starting from Eqs. \eqref{eq:intro_lindblad_equation} and \eqref{eq:lindblad_operator}
we can now construct an unravelling consistent with the quantum state
diffusion (QSD) approach \cite{gisin1992quantum,gisin1993quantum,strunz1996linear,brun2000continuous,percival1998quantum}.
Using the set of Kraus operators \cite{jacobs2014quantum,clarke2021irreversibility,clarke2022entropy}
\begin{equation}
M_{\pm k}\propto\left(\mathbb{I}-iH_{sys}dt-\frac{1}{2}L_{k}^{\dagger}L_{k}dt\pm L_{k}\sqrt{dt}\right),\label{eq:kraus_operator}
\end{equation}
consistent with the development in Section \ref{sec:intro_fokker_planck_equation},
and substituting Eq. \eqref{eq:kraus_operator} into Eq. \eqref{eq:change_kraus_operators}
we obtain (see Refs. \cite{jacobs2014quantum,clarke2022entropy}):

\begin{equation}
d\rho=-i\left[H_{sys},\rho\right]dt+\sum_{k}\left[\left(L_{k}\rho L_{k}^{\dagger}-\frac{1}{2}\left\{ L_{k}^{\dagger}L_{k},\rho\right\} \right)dt+\left(\rho L_{k}^{\dagger}+L_{k}\rho-{\rm Tr}\left[\rho\left(L_{k}+L_{k}^{\dagger}\right)\right]\rho\right)dW_{k}\right],\label{eq:unravelled_lindblad_eq}
\end{equation}
where we use $\rho$ to denote the physical reduced density matrix
of the system that evolves stochastically, and where environmental
noise is represented by a set of independent Wiener increments $dW_{k}$.
Notice that upon averaging over the Wiener noises this evolution equation
corresponds to Eq. \eqref{eq:intro_lindblad_equation} and hence the
density matrix $\bar{\rho}$ corresponds to the noise averaged density
matrix $\rho$. Eq. \eqref{eq:unravelled_lindblad_eq} finally provides
us with the dynamics and the means by which to determine the irreversibility
of system behaviour in various cases.

\subsubsection{Equations of motion for components of the coherence vector}

With the particular environmental couplings defined in Eq. \eqref{eq:couplings_environments}
($k=x,y,z$), and with $\gamma_{0}=1$, we obtain Lindblad operators:

\begin{equation}
L_{x}=\lambda\left(\sigma_{x}-i\frac{\beta\epsilon}{2}\sigma_{y}\right),\hspace{2em}L_{y}=\lambda\left(\sigma_{y}+i\frac{\beta\epsilon}{2}\sigma_{x}\right),\hspace{2em}L_{z}=\lambda\gamma\sigma_{z}.
\end{equation}
Instead of simulating the dynamics of elements of the density matrix
directly, we will evolve the coherence vector defined by the three
components $r_{i}=\text{Tr}(\sigma_{i}\rho)$. This is consistent
with a representation of the density matrix in the form
\begin{equation}
\rho=\frac{1}{2}\left(\mathbb{I}+\boldsymbol{r}\cdot\boldsymbol{\sigma}\right).\label{eq:coherence vector rep}
\end{equation}
We obtain from Eq. \eqref{eq:unravelled_lindblad_eq} the following
equations:

\begin{equation}
dr_{x}=-2\epsilon r_{y}dt-2\lambda^{2}\left(1+\gamma^{2}\right)r_{x}dt+\lambda\left(\beta\epsilon r_{z}+2\left(1-r_{x}^{2}\right)\right)dW_{x}-2\lambda r_{x}r_{y}dW_{y}-2\gamma\lambda r_{x}r_{z}dW_{z}\label{eq:dr_x_general}
\end{equation}

\begin{equation}
dr_{y}=2\epsilon r_{x}dt-2\lambda^{2}\left(1+\gamma^{2}\right)r_{y}dt-2\lambda r_{x}r_{y}dW_{x}+\lambda\left(\beta\epsilon r_{z}+2\left(1-r_{y}^{2}\right)\right)dW_{y}-2\gamma\lambda r_{y}r_{z}dW_{z}\label{eq:dr_y_general}
\end{equation}

\begin{equation}
dr_{z}=-4\lambda^{2}\left(\beta\epsilon+r_{z}\right)dt-\lambda r_{x}\left(\beta\epsilon+2r_{z}\right)dW_{x}-\lambda r_{y}\left(\beta\epsilon+2r_{z}\right)dW_{y}+2\gamma\lambda\left(1-r_{z}^{2}\right)dW_{z}.\label{eq:dr_z_general}
\end{equation}
Notice how important it is to consider the \emph{stochastic} evolution
of the coherence vector rather than the averaged behaviour. For the
latter situation, described by $d\langle r_{x}\rangle=-2\epsilon\langle r_{y}\rangle dt-2\lambda^{2}\left(1+\gamma^{2}\right)\langle r_{x}\rangle dt$,
etc, any initial state results in $\langle r_{x}\rangle\rightarrow0$,
$\langle r_{y}\rangle\rightarrow0$ and $\langle r_{z}\rangle\rightarrow-\beta\epsilon$
at long times and hence is not disturbed by the evolution of $\gamma$
away from unity. What does change in these circumstances is the pdf
of the system variables, and hence higher moments of the components
of the coherence vector. The noise-averaged Lindblad equation for
$\bar{\rho}$ cannot capture the stochastic selection of an eigenstate
under measurement.

\subsubsection{Equations of motion in cylindrical coordinates \label{sec:Cylindrical-coordinate-equations}}

Next we derive the equations of motion for the convenient set of coordinates
$(r^{2},r_{z},\phi)$, where 
\begin{equation}
r^{2}=r_{x}^{2}+r_{y}^{2}+r_{z}^{2}\,\,\,\text{and}\,\,\,\phi=\arctan\left(\frac{r_{y}}{r_{x}}\right).\label{eq:cylindrical_coords}
\end{equation}
Using Itô's Lemma, see Appendix \ref{subsec:Cylindrical-coordinate-equations},
we can obtain:
\begin{equation}
\begin{aligned}dr^{2} & =4\lambda^{2}\left(r^{2}-1\right)\left(\gamma^{2}r_{z}^{2}-\gamma^{2}+r^{2}-r_{z}^{2}-2\right)dt+4\lambda\sqrt{r^{2}-r_{z}^{2}}\cos\phi\left(1-r^{2}\right)dW_{x}\\
 & +4\lambda\sqrt{r^{2}-r_{z}^{2}}\sin\phi\left(1-r^{2}\right)dW_{y}+4\lambda\gamma r_{z}\left(1-r^{2}\right)dW_{z}\,,
\end{aligned}
\label{eq:dr_squared}
\end{equation}

\begin{align}
dr_{z} & =-4\lambda^{2}\left(\beta\epsilon+r_{z}\right)dt-\lambda\left(\beta\epsilon+2r_{z}\right)\sqrt{r^{2}-r_{z}^{2}}\cos\phi\,dW_{x}\nonumber \\
 & -\lambda\left(\beta\epsilon+2r_{z}\right)\sqrt{r^{2}-r_{z}^{2}}\sin\phi\,dW_{y}+2\gamma\lambda\left(1-r_{z}^{2}\right)\,dW_{z}\,,
\end{align}
and

\begin{equation}
d\phi=2\epsilon dt-\lambda\frac{\left(\beta\epsilon r_{z}+2\right)\sin\phi}{\sqrt{r^{2}-r_{z}^{2}}}dW_{x}+\lambda\frac{\left(\beta\epsilon r_{z}+2\right)\cos\phi}{\sqrt{r^{2}-r_{z}^{2}}}dW_{y}\,.
\end{equation}
Due to the coordinate singularities, suitable care needs to be taken
with regard to initial state choices and the dynamics themselves.

\subsubsection{Initial state simplification}

We see from Eq. \eqref{eq:dr_squared} that $r=1$ is a fixed point
of the dynamics. Defining a pure state by the condition ${\rm Tr}(\rho^{2})=1$,
it can be shown that $r=1$ corresponds to a pure state by inserting
$\rho$ in its coherence vector representation \ref{eq:coherence vector rep}.
In order to reduce the computational difficulty of solving the equations,
we set $r=1$ and consider

\begin{align}
dr_{z} & =-4\lambda^{2}\left(\beta\epsilon+r_{z}\right)dt-\lambda\left(\beta\epsilon+2r_{z}\right)\sqrt{1-r_{z}^{2}}\cos\phi\,dW_{x}\label{eq:rz_eq_simplified}\\
 & -\lambda\left(\beta\epsilon+2r_{z}\right)\sqrt{1-r_{z}^{2}}\sin\phi\,dW_{y}+2\gamma\lambda\left(1-r_{z}^{2}\right)\,dW_{z}\nonumber 
\end{align}

\begin{equation}
d\phi=2\epsilon dt-\lambda\frac{\left(\beta\epsilon r_{z}+2\right)\sin\phi}{\sqrt{1-r_{z}^{2}}}dW_{x}+\lambda\frac{\left(\beta\epsilon r_{z}+2\right)\cos\phi}{\sqrt{1-r_{z}^{2}}}dW_{y}.\label{eq:phi_eq_simplified}
\end{equation}
Hence, the coherence vector describing our system will always lie
on the surface of the Bloch sphere, and the pdf will depend only on
two variables: $r_{z}$ and $\phi$. These equations form the basis
of our simulations with corresponding results to be presented in Section
\ref{sec:Results}.

\subsection{Fokker-Planck equation\label{sec:fokker_planck_system}}

\begin{figure}[H]
\begin{centering}
\includegraphics[width=0.65\linewidth]{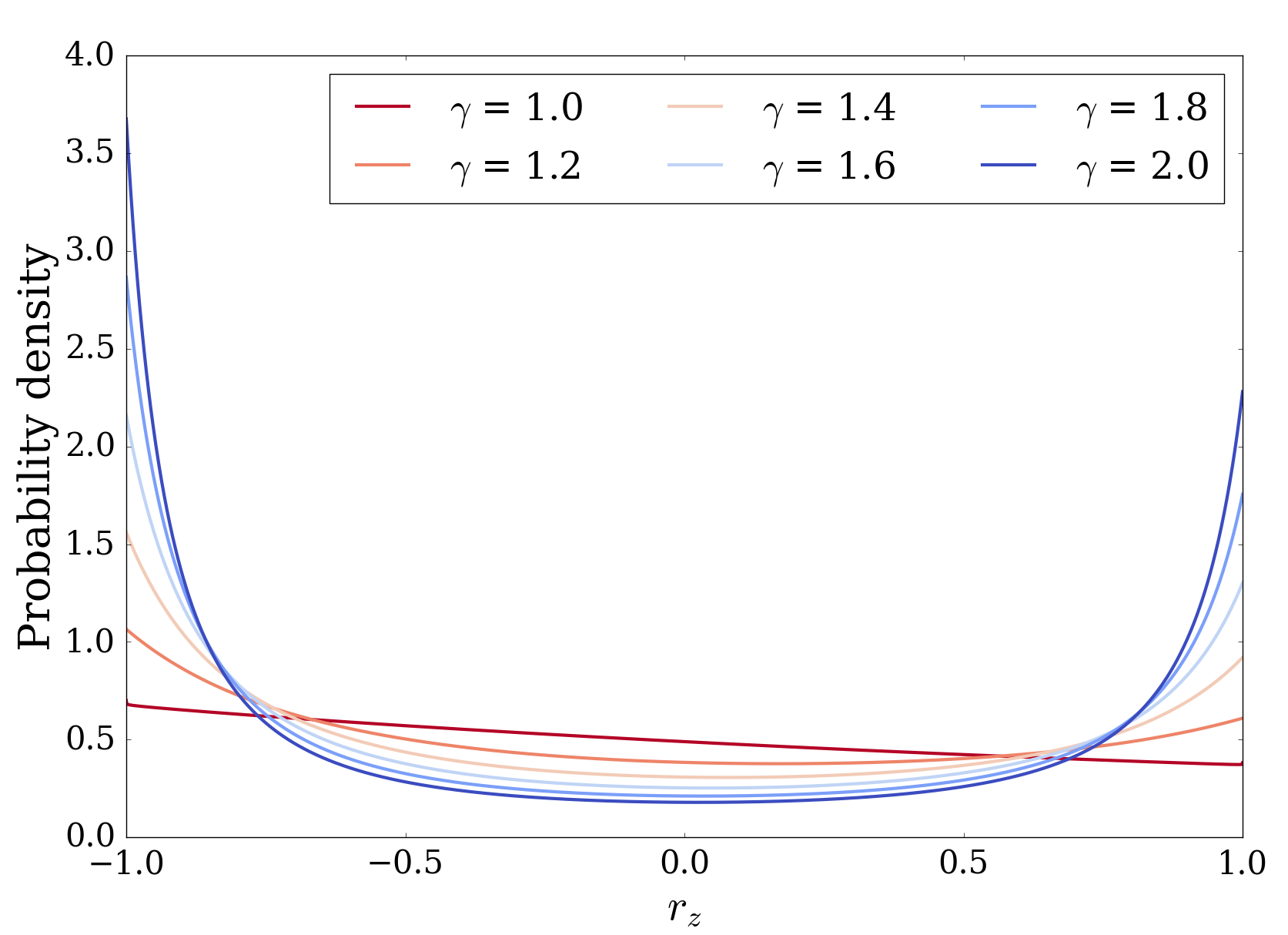}
\par\end{centering}
\caption{Stationary pdf solutions of the Fokker-Planck Eq. \eqref{eq:fokker_planck_rz}
for different values of $\gamma$. Parameters used are $\beta=0.1$
and $\epsilon=1$.}
\label{fig:fp_equation}
\end{figure}

The definition of the entropy production in Eq. \eqref{eq:entropy_production}
requires an evolving pdf for the system coordinates over time, and
hence we need to derive and solve the associated Fokker-Planck equation.
To do so, we need expressions for the vector $\boldsymbol{A}$ and
matrices $\boldsymbol{B}$ and \textbf{$\boldsymbol{D}=\frac{1}{2}\boldsymbol{B}\boldsymbol{B}^{T}$},
obtained by comparing their definitions in Eq. \eqref{eq:general_SDE}
with the equations of motion \eqref{eq:rz_eq_simplified} and \eqref{eq:phi_eq_simplified}.
We get
\begin{equation}
\boldsymbol{A}=\left(\begin{matrix}A_{z}\\
A_{\phi}
\end{matrix}\right)=\left(\begin{matrix}-4\lambda^{2}(\beta\epsilon+r_{z})\\
2\epsilon
\end{matrix}\right),\label{eq:mu_array_rz_phi}
\end{equation}

\begin{equation}
\boldsymbol{B}=\lambda\left(\begin{matrix}-(\beta\epsilon+2r_{z})\sqrt{1-r_{z}^{2}}\cos\phi & -(\beta\epsilon+2r_{z})\sqrt{1-r_{z}^{2}}\sin\phi & 2\gamma(1-r_{z}^{2})\\
-(\beta\epsilon r_{z}+2)\frac{\sin\phi}{\sqrt{1-r_{z}^{2}}} & (\beta\epsilon r_{z}+2)\frac{\cos\phi}{\sqrt{1-r_{z}^{2}}} & 0
\end{matrix}\right),\label{eq:B_matrix_rz_phi}
\end{equation}
and hence

\begin{equation}
\boldsymbol{D}=\left(\begin{array}{cc}
D_{zz} & 0\\
0 & D_{\phi\phi}
\end{array}\right)\label{eq:D_matrix_rz_phi}
\end{equation}
with 
\begin{equation}
D_{zz}=2\lambda^{2}\left(1-r_{z}^{2}\right)\left(\left(\frac{\beta\epsilon}{2}\right)^{2}+\beta\epsilon r_{z}+(1-\gamma^{2})r_{z}^{2}+\gamma^{2}\right)\label{eq:D_zz}
\end{equation}
and

\begin{equation}
D_{\phi\phi}=2\lambda^{2}\frac{\left(\frac{\beta\epsilon r_{z}}{2}\right)^{2}+\beta\epsilon r_{z}+1}{1-r_{z}^{2}}\,.\label{eq:Dphiphi}
\end{equation}
There exists no $\phi$ dependence in either $\boldsymbol{A}$ or
$\boldsymbol{D}$, and the latter matrix is indeed diagonal, as anticipated.
With these expressions, the Fokker-Planck equation for the pdf $p(r_{z},\phi,t)$
becomes:

\begin{align}
\frac{\partial}{\partial t}p(r_{z},\phi,t) & =\frac{\partial}{\partial r_{z}}\left[-A_{z}p(r_{z},\phi,t)+\frac{\partial}{\partial r_{z}}\left(D_{zz}p(r_{z},\phi,t)\right)\right]+\frac{\partial}{\partial\phi}\left[-A_{\phi}p(r_{z},\phi,t)+\frac{\partial}{\partial\phi}\left(D_{\phi\phi}p(r_{z},\phi,t)\right)\right]\label{eq:fokker_planck_with_phi}\\
 & =\frac{\partial}{\partial r_{z}}\left[\frac{\lambda^{2}}{2}\left(-2\beta^{2}\epsilon^{2}r_{z}+12\beta\epsilon\left(1-r_{z}^{2}\right)+16(1-\gamma^{2})r_{z}\left(1-r_{z}^{2}\right)\right)p(r_{z},\phi,t)\right.\nonumber \\
 & \qquad\qquad\left.+2\lambda^{2}\left(1-r_{z}^{2}\right)\left(\left(\frac{\beta\epsilon}{2}\right)^{2}+\beta\epsilon r_{z}+(1-\gamma^{2})r_{z}^{2}+\gamma^{2}\right)\frac{\partial}{\partial r_{z}}p(r_{z},\phi,t)\right]\nonumber \\
 & \qquad\qquad+\frac{\partial}{\partial\phi}\left[-2\epsilon p(r_{z},\phi,t)+2\lambda^{2}\frac{\left(\frac{\beta\epsilon r_{z}}{2}\right)^{2}+\beta\epsilon r_{z}+1}{1-r_{z}^{2}}\frac{\partial}{\partial\phi}p(r_{z},\phi,t)\right].\nonumber 
\end{align}
The Fokker-Planck equation can be simplified further if we assume
the initial pdf to be independent of $\phi$. As there is no explicit
$\phi$ dependence in Eq. \eqref{eq:fokker_planck_with_phi}, if the
initial pdf is $\phi$ independent, it will always remain so. This
leads us to the simplification $p(r_{z},\phi,t)\rightarrow p(r_{z},t)$,
and a more succinct equation:
\begin{align}
\frac{\partial}{\partial t}p(r_{z},t) & =\frac{\partial}{\partial r_{z}}\left[\frac{\lambda^{2}}{2}\left(-2\beta^{2}\epsilon^{2}r_{z}+12\beta\epsilon\left(1-r_{z}^{2}\right)+16(1-\gamma^{2})r_{z}\left(1-r_{z}^{2}\right)\right)p(r_{z},t)\right.\label{eq:fokker_planck_rz}\\
 & \left.\qquad\qquad+2\lambda^{2}\left(1-r_{z}^{2}\right)\left(\left(\frac{\beta\epsilon}{2}\right)^{2}+\beta\epsilon r_{z}+(1-\gamma^{2})r_{z}^{2}+\gamma^{2}\right)\frac{\partial}{\partial r_{z}}p(r_{z},t)\right].\nonumber 
\end{align}
The stationary solution $p_{st}(r_{z})$ of the Fokker-Planck equation
corresponding to setting $\partial p(r_{z},t)/\partial t=0$ can be
obtained analytically using tools such as Mathematica. For $\gamma\neq1$,
we have:

\begin{align}
p_{st}(r_{z})_{\gamma\neq1} & =\frac{1}{N_{\gamma}}(1-r_{z})^{a}(1+r_{z})^{b}\left(\beta^{2}\epsilon^{2}+4\beta\epsilon r_{z}+4\gamma^{2}+4(1-\gamma^{2})\right)^{c}\label{eq:fp_eq_general_gamma}\\
 & \qquad\qquad\qquad\times\exp\left[\frac{8\beta\epsilon\left(\beta^{2}\epsilon^{2}(1+2\gamma^{2})-4\right){\rm arctanh}\left[\frac{-\beta\epsilon-2r_{z}(1-\gamma^{2})}{\gamma\sqrt{4\gamma^{2}-4+\beta^{2}\epsilon^{2}}}\right]}{(4-\beta^{2}\epsilon^{2})^{2}\gamma\sqrt{4\gamma^{2}-4+\beta^{2}\epsilon^{2}}}\right],\nonumber 
\end{align}
while for $\gamma=1$ we have

\begin{equation}
p_{st}(r_{z})_{\gamma=1}=\frac{1}{N_{1}}(1-r_{z})^{a}(1+r_{z})^{b}(\beta^{2}\epsilon^{2}+4\beta\epsilon r_{z}+4)^{d},\label{eq:fp_eq_gamma1}
\end{equation}
where $N_{\gamma}$ is a $\gamma$ dependent normalisation factor,
and the exponents are given by 
\begin{equation}
a=-\left(\frac{\beta\epsilon}{2+\beta\epsilon}\right)^{2},\,\,\,b=-\left(\frac{\beta\epsilon}{2-\beta\epsilon}\right)^{2},\,\,\,c=-1+\frac{4(-4+3\beta^{2}\epsilon^{2})}{(-4+\beta^{2}\epsilon^{2})^{2}},\,\,\,d=-1+\frac{8(-4+3\beta^{2}\epsilon^{2})}{(-4+\beta^{2}\epsilon^{2})^{2}},
\end{equation}
where we note that $c\neq d$. The stationary pdfs in Eqs. \eqref{eq:fp_eq_general_gamma}
and \eqref{eq:fp_eq_gamma1} are shown in Figure \ref{fig:fp_equation}
for a number of values of $\gamma$.

We can see that the coupling strength $\gamma$ has a very significant
effect on the stationary probability density of the system coordinate
$r_{z}$. Increasing $\gamma$ concentrates the pdf closer to the
boundaries and in an asymmetric manner due to the system Hamiltonian
bias $\epsilon$. By evolving $\gamma$ from unity to a larger value,
we can capture the effect of a measurement of the system energy, as
this would cause individual trajectories to dwell in the vicinity
of the eigenstates of $H_{sys}$ at $r_{z}=\pm1$, or equivalently
at system energies $\text{Tr}\left(H_{sys}\rho\right)=\epsilon r_{z}$
of $\pm\epsilon$. 

\subsection{Average system entropy production and boundary contributions\label{sec:boundary_term_effects}}

Using its definition in Eq. \eqref{eq:stochastic_system_entropy},
the incremental stochastic system entropy production over each time
step $dt$ can be calculated from the time-dependent pdf $p(r_{z},t)$
obtained from the Fokker-Planck equation \eqref{eq:fokker_planck_rz}.
We then average over many realisations of the behaviour. However,
we have noticed in practice that combining the average system entropy
production with the average environmental entropy production Eq. \eqref{eq:stochastic_environmental_entropy},
obtained numerically for time independent environmental coupling,
does not yield a vanishing total entropy production that we would
expect for a system in a stationary state with zero probability current.
We have therefore investigated this situation further. The incremental
system entropy production averaged over $r_{z}$ and the noises for
our system may be written as

\begin{align}
d\langle\langle\Delta s_{sys}\rangle\rangle & =\int_{-1}^{1}d\langle\Delta s_{sys}\rangle\,p(r_{z},t)dr_{z},\label{incremental mean sys entropy}
\end{align}
where $d\langle\Delta s_{sys}\rangle$ is the noise averaged increment
in system entropy production, which is a function of $r_{z}$ and
$t$. By integrating by parts twice and retaining the boundary terms,
see Appendix \ref{subsec:Averaged-system-entropy} for details, we
obtain:
\begin{align}
d\langle\langle\Delta s_{sys}\rangle\rangle & =dS_{G}-\left[J_{z}\ln p(r_{z},t)\right]_{-1}^{1}dt-\left[D_{zz}\frac{\partial p(r_{z},t)}{\partial r_{z}}\right]_{-1}^{1}dt,\label{eq:boundary_term}
\end{align}
where $dS_{G}$ is the increment in the Gibbs entropy of the system,
shown in Eq. \eqref{eq:averaged_system_entropy_production}, and $J_{z}$
is the $r_{z}$-component of the probability current defined in Appendix
\ref{subsec:Averaged-system-entropy}.

In Eq. \eqref{eq:averaged_system_entropy_production} we noted that
$d\langle\langle\Delta s_{sys}\rangle\rangle=dS_{G}$ held only under
certain conditions, now shown to be where the boundary terms in Eq.
\eqref{eq:boundary_term} vanish. For example, the current $J_{z}$
and often the pdf and its derivatives vanish at the boundaries of
the parameter space, and diffusion coefficients can also go to zero.
For our system, although $D_{zz}=0$ at $r_{z}=\pm1$, it can be shown
that the gradient $\partial p_{st}(r_{z},t)/\partial r_{z}$ is singular
at the boundaries for the stationary solution of the Fokker-Planck
equation, and that by implication this applies to nonstationary situations
as well. There is therefore a need to correct the usual result $d\langle\langle\Delta s_{sys}\rangle\rangle=dS_{G}$.

In contrast to this analysis, we observe a mean system entropy production
consistent with zero when our system is in a stationary state, which
suggests that the numerically obtained time dependent pdf is not sufficiently
accurate near the boundaries, or the sampling of these regions by
the generated trajectories is too limited, to produce the appropriate
additional contributions. Our approach, therefore, is to estimate
the boundary terms analytically for the stationary pdf and to add
them to the numerical results.

Using the stationary pdfs given in Eqs. \eqref{eq:fp_eq_gamma1} and
\eqref{eq:fp_eq_general_gamma}, together with the diffusion coefficient
$D_{zz}$ given in Eq. \eqref{eq:D_matrix_rz_phi}, it may be shown
that the boundary correction with $J_{z}$ in Eq. \eqref{eq:boundary_term}
vanishes at equilibrium, and so does $dS_{G}.$ This allows us to
write the mean stationary system entropy production increments for
general $\gamma$ as 
\begin{equation}
d\langle\langle\Delta s_{sys}\rangle\rangle_{st}=-\left[D_{zz}\frac{\partial p_{st}(r_{z})_{\gamma}}{\partial r_{z}}\right]_{-1}^{1}dt,\label{eq:boundary correction}
\end{equation}
which does not vanish. In fact, it is possible to obtain a simple
analytical expression for $\gamma=1$ to lowest order in $\beta\epsilon$:
\begin{equation}
d\langle\langle\Delta s_{sys}\rangle\rangle_{st}^{\gamma=1}=-\lambda^{2}(\beta\epsilon)^{2}dt.\label{eq:boundary_correction_gamma1}
\end{equation}
This shows that this boundary correction can be obtained exactly analytically
for $\gamma=1$, though it becomes harder for $\gamma\neq1$. In the
latter case we have a more complicated form of $p_{st}(r_{z})$, and
hence it is necessary to expand it analytically in terms of small
$\beta\epsilon$ before calculating the boundary correction.

\subsection{Average environmental entropy production}

With the choice of the couplings and Hamiltonian, we can calculate
the environmental entropy production from Eq. \eqref{eq:stochastic_system_entropy}.
We need to determine \textbf{$\boldsymbol{A}^{irr}$} and $\boldsymbol{A}^{rev}$
from Eqs. \eqref{eq:rz_eq_simplified} and \eqref{eq:phi_eq_simplified}.
This is done by understanding how the components of the coherence
vector and the various contributions to $\boldsymbol{A}$ transform
under time reversal. For a system without spin degrees of freedom,
the time reversal operator is given by $\Theta=K$ , where $K$ is
the operation of complex conjugation \cite{sachs1987physics} and
the time reversal operation on the Pauli matrices produces

\begin{equation}
\Theta\sigma_{z}\Theta^{-1}=\sigma_{z}\qquad\Theta\sigma_{y}\Theta^{-1}=-\sigma_{y}\qquad\Theta\sigma_{x}\Theta^{-1}=\sigma_{x}.
\end{equation}
Thus $r_{x}$ and $r_{z}$ are even under time reversal, and $r_{y}$
is odd, and hence $\phi$ is also odd, and we can separate the coefficients
of the $dt$ terms in Eqs. \eqref{eq:rz_eq_simplified} and \eqref{eq:phi_eq_simplified}
into their $\boldsymbol{A}^{irr}$ and $\boldsymbol{A}^{rev}$components.
We can write

\begin{equation}
\boldsymbol{A}^{irr}=\left(\begin{matrix}-4\lambda^{2}\left(\beta\epsilon+r_{z}\right)\\
0
\end{matrix}\right),\hspace{2em}\boldsymbol{A}^{rev}=\left(\begin{matrix}0\\
2\epsilon
\end{matrix}\right)\,,
\end{equation}
and obtain an expression for the environmental entropy production
using Eq. \eqref{eq:stochastic_environmental_entropy}, since we already
have the form of $\boldsymbol{D}$ from Eq. \eqref{eq:D_matrix_rz_phi}.

\section{Simulations and Results\label{sec:Results}}

In this section we describe two protocols designed to represent the
dynamics of connecting the system to and disconnecting it from a measuring
device, respectively, and we compute the associated stochastic entropy
production for each.

\subsection{The computational procedure}

In order to demonstrate adherence to the detailed fluctuation theorem
we consider two simple protocols that are a time reversal of each
other:
\begin{itemize}
\item Connecting a measuring device (protocol M): begin at $t=0$ with the
system thermalised for $\gamma=1$ and the initial pdf defined as
in Figure \ref{fig:fp_equation}; for $t>0$ perform simulations of
Eqs. \eqref{eq:rz_eq_simplified} and \eqref{eq:phi_eq_simplified}
using $\gamma=2$;
\item Disconnecting a measuring device (protocol $\overline{{\rm M}}$):
begin at $t=0$ with the system thermalised for $\gamma=2$ and the
initial pdf defined as in Figure \ref{fig:fp_equation}; then, for
$t>0$ perform simulations of Eqs. \eqref{eq:rz_eq_simplified} and
\eqref{eq:phi_eq_simplified} using $\gamma=1$.
\end{itemize}
The thermalisation of the system is defined by the pdf given by the
stationary Fokker-Planck equation as plotted in Figure \ref{fig:fp_equation}.
For both protocols, the initial value of $r_{z}$ for a trajectory
is randomly sampled from the appropriate pdf $p_{st}(r_{z})$ corresponding
to either $\gamma=1$ or $\gamma=2$, while the value of $\phi$ is
randomly sampled from a uniform distribution within the range $[0,2\pi)$.
Whenever we specify a particular value of $\gamma$, we will be referring
to the value used for the dynamics (i.e., for $t>0$), unless we explicitly
mention the thermalisation condition ($t=0$). For both protocols
several parameters of the system are kept constant: the environmental
inverse temperature $\beta=0.1$, the Hamiltonian parameter $\epsilon=1$,
the proportionality constant in the spectral density $\alpha=0.01$,
and $\lambda=\sqrt{0.2}$.

In order to calculate the system entropy production, we solve the
Fokker-Planck equation \eqref{eq:fokker_planck_rz} for $r_{z}$ and
observe the behaviour of the system going from a stationary state
at a particular value of $\gamma$ to another stationary state defined
by a different $\gamma$. The protocols will drive the system between
two states obtained from the Fokker-Planck equation at different $\gamma$
values. The evolution from one stationary pdf to the other occurs
over a timescale of order $t=1$, for both protocols.

In summary, the computational procedure is as follows:
\begin{enumerate}
\item Obtain the solution $p(r_{z},t)$ of the Fokker-Planck equation \eqref{eq:fokker_planck_rz}
using the appropriate stationary pdf as the initial condition (at
$t=0$) for the trajectories; in both cases the boundary conditions
at $r_{z}=\pm1$ are chosen corresponding to a zero probability current
$J_{z}$ defined in Appendix \ref{subsec:Averaged-system-entropy}.
Note that each pdf is obtained on a grid of $r_{z}$ values.
\item Start a loop over the noises (independently for each protocol):
\begin{enumerate}
\item generate noises $dW_{x}$, $dW_{y}$ and $dW_{z}$ for each time step
$dt$;
\item run the stochastic dynamics for $r_{z}(t)$ and $\phi(t)$ for both
protocols using Eqs. \eqref{eq:rz_eq_simplified} and \eqref{eq:phi_eq_simplified}
based on initial values sampled from the corresponding $\gamma$-dependent
stationary pdf;
\item for each time increment $dt$ and using obtained values of $r_{z}(t)$
and $\phi(t)$, calculate the incremental contributions to the environmental,
$d\Delta s_{env}$, and system, $d\Delta s_{sys}=-d\left(\ln\,p(r_{z},t)\right)$,
stochastic entropy productions via respectively, Eq. \eqref{eq:stochastic_environmental_entropy}
and the first line of Eq. \eqref{eq:stochastic_system_entropy}; in
the latter case the increment of the logarithm of the pdf is calculated
as a difference between its values at two consecutive times. The values
of the pdf $p(r_{z},t)$ for these times at the required value of
$r_{z}$ are obtained by linearly interpolating each pdf between the
two nearest $r_{z}$ values on the grid;
\item the incremental total stochastic entropy production, at each time
step $dt$, is calculated as a sum, $d\Delta s_{tot}=d\Delta s_{env}+d\Delta s_{sys}$;
accumulate these contributions over an entire trajectory;
\item go back to step 2(a) to run another trajectory; the calculation is
repeated the required number of times using different manifestations
of the noises. The stochastic entropy production is averaged over
the trajectories. Each protocol will have its corresponding independent
ensemble of entropy productions.
\end{enumerate}
\item Once the averaged values of the entropy production increments are
obtained from a large ensemble of trajectories, the boundary term
from Eq. \eqref{eq:boundary_term} is added to the ensemble average
$d\left\langle \left\langle \Delta s_{tot}\right\rangle \right\rangle $
as 
\begin{equation}
d\langle\langle\Delta s_{tot}\rangle\rangle=d\langle\langle\Delta s_{env}\rangle\rangle+d\langle\langle\Delta s_{sys}\rangle\rangle-\left[D_{zz}\frac{\partial p_{st}(r_{z})}{\partial r_{z}}\right]_{-1}^{1}.
\end{equation}
 for the specific value of $\gamma$. This yields the final increment
of the mean total entropy production $d\langle\langle\Delta s_{tot}\rangle\rangle$
for the given time step $dt$, separately for the two protocols.
\end{enumerate}

\subsection{Dynamics of $r_{z}$ and $\phi$}

\begin{figure}[H]
\centering{}\centering \includegraphics[width=0.4\linewidth]{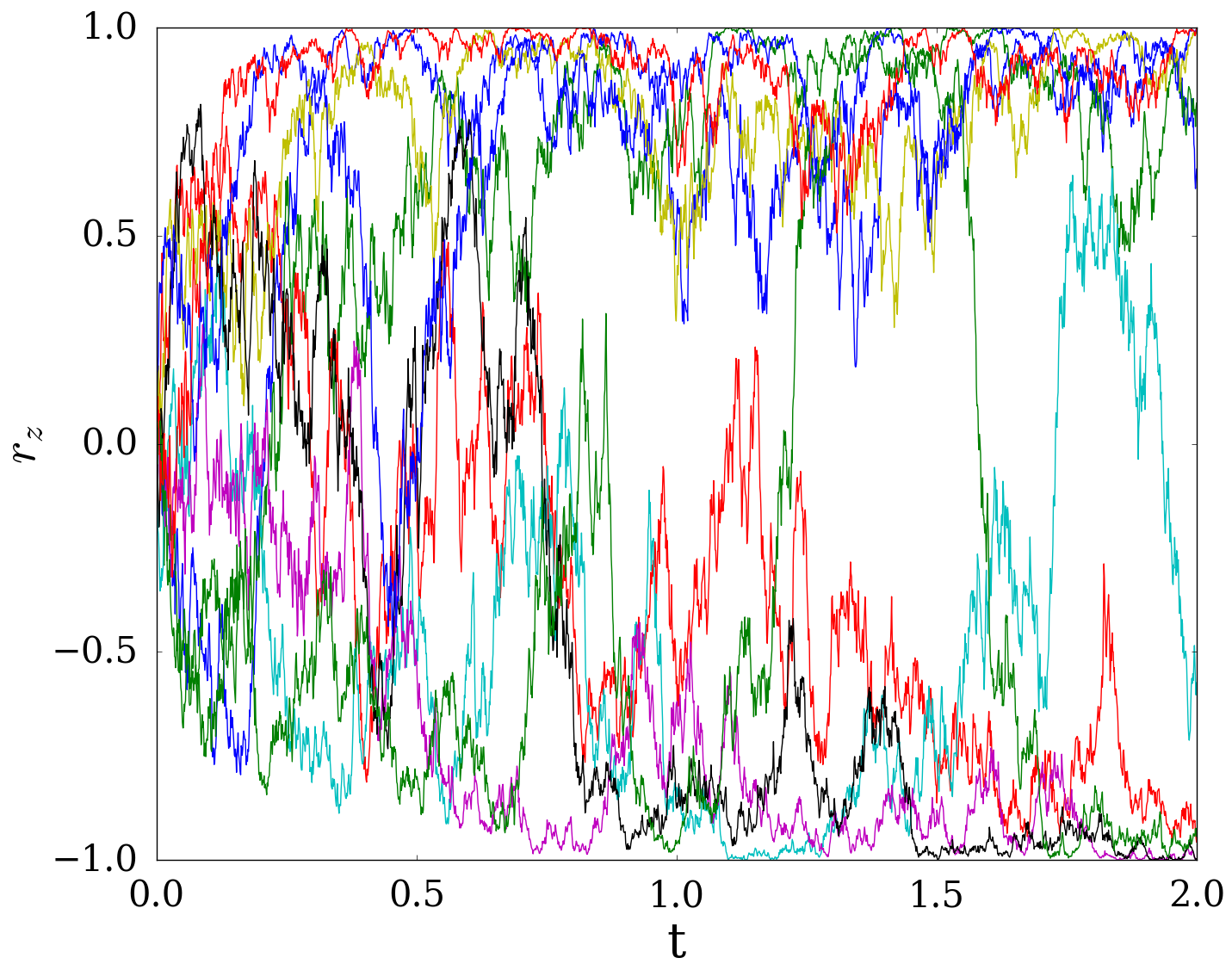}
\includegraphics[width=0.45\linewidth]{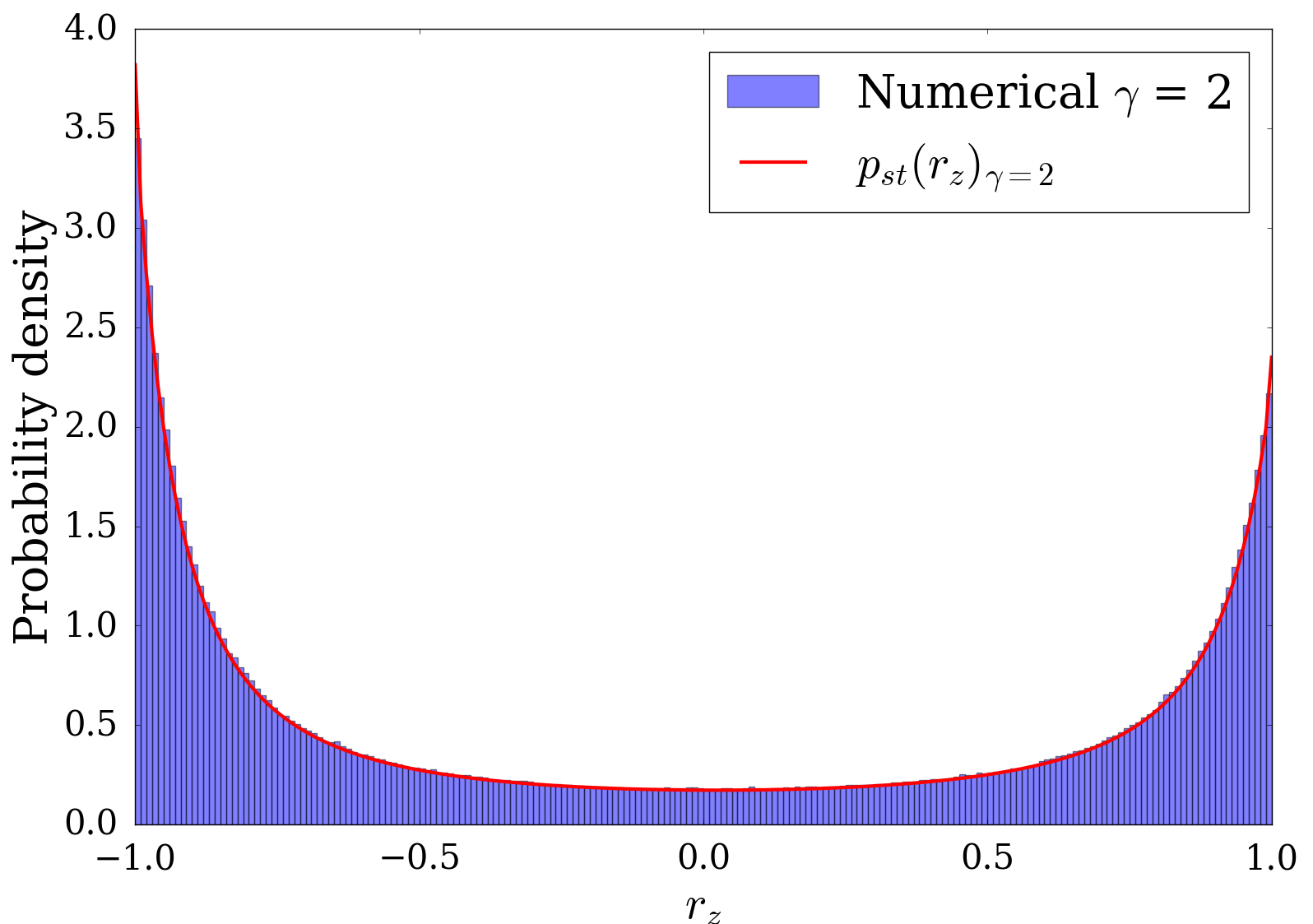}
\caption{Left panel: a set of 10 random trajectories that were initialised
at $r_{z}=0$ for protocol M ($\gamma=2$), displaying the drifting
behaviour of the system towards the eigenstates driven by the measuring
device. Right panel: a normalised histogram for protocol M based on
one million individual trajectories at $t=2,$ and compared with $p_{st}(r_{z})_{\gamma=2}$
of Eq. \eqref{eq:fp_eq_general_gamma} obtained by solving the Fokker-Planck
equation.}
\label{fig:trajectories_eigenstates}
\end{figure}

\begin{figure}[H]
\centering \includegraphics[width=0.4\linewidth]{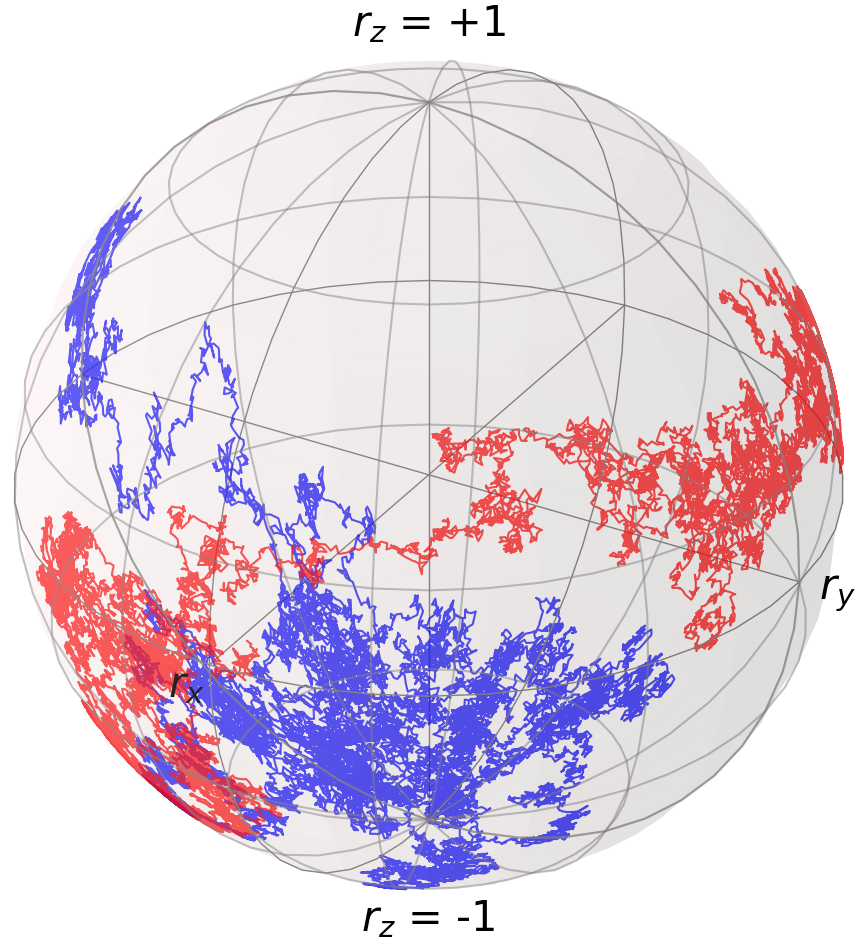}
\includegraphics[width=0.48\linewidth]{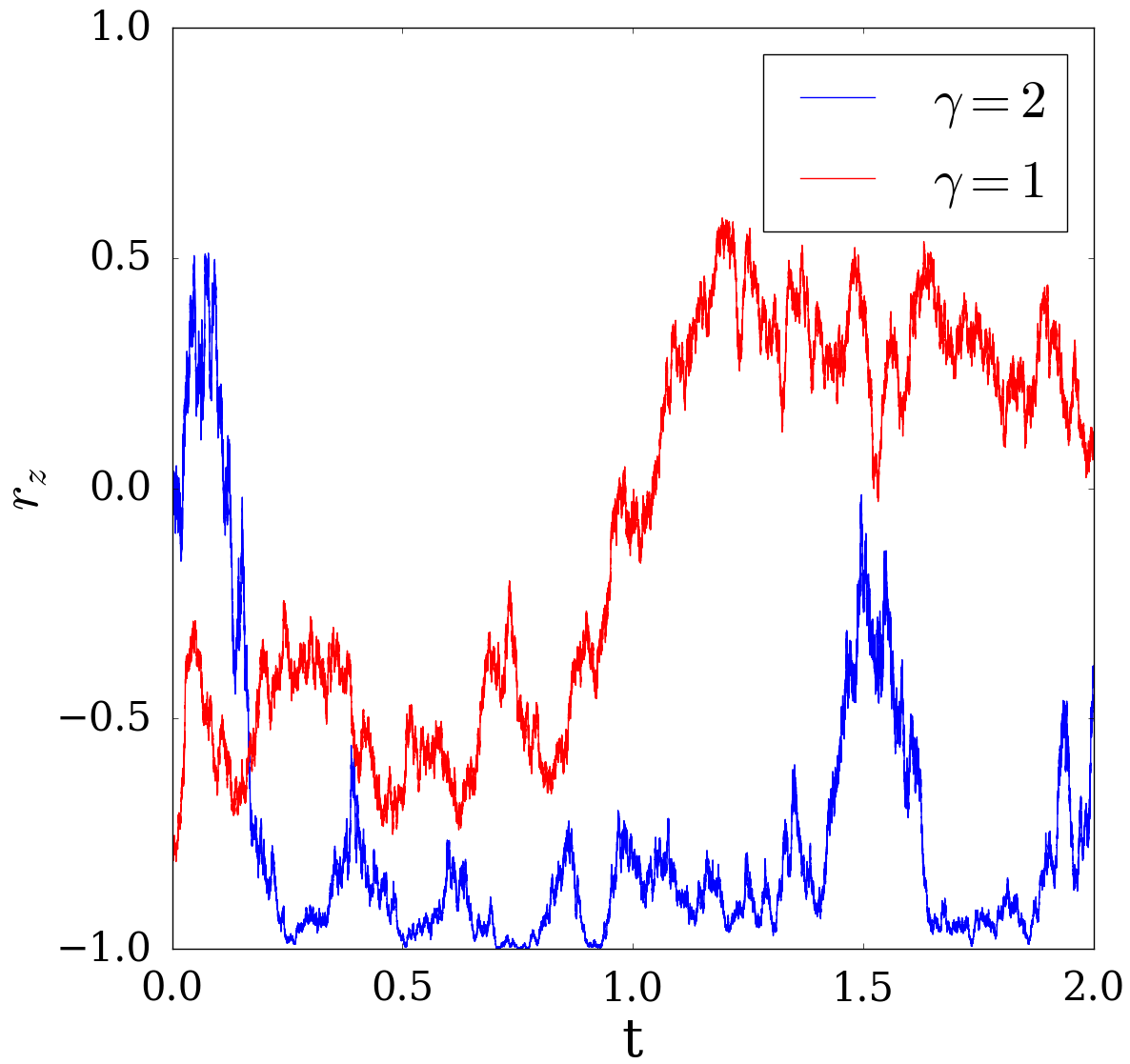}
\caption{Example individual trajectories for the two protocols ($\gamma=2$,
blue; $\gamma=1$, red) , exploring the Bloch sphere. Parameters used
are: $\beta=0.1$, $\epsilon=1$, $t_{max}=2$, $dt=10^{-5}$, and
$\alpha=0.01$. Initial states for the protocols were randomly sampled
from the stationary solutions to the Fokker-Planck equation, for $\gamma=1$
and $\gamma=2$, respectively.}
\label{fig:trajectory_bloch_sphere}
\end{figure}

The thermal state $\bar{\rho}_{eq}$ of the system is approximately
proportional to $e^{-\beta H_{sys}}$ for weak system-environment
coupling. In the derivation of Eq. \eqref{eq:unravelled_lindblad_eq}
we have assumed $\beta H_{sys}\ll1,$ so we can write $\bar{\rho}_{eq}\approx\frac{1}{2}(\mathbb{I}-\beta H_{sys})$
and hence obtain the average of $r_{z}$ as $\bar{r}_{z}={\rm Tr}\left(\sigma_{z}\bar{\rho}_{eq}\right)=-\beta\epsilon\approx-0.1$
for our parameters. We find that our numerics supports this: by running
one million realisations of the dynamics for protocol M (representing
the connection of the measuring device), we observe that the averaged
results of $r_{z}$ do indeed remain constant throughout the simulation.
Furthermore, the dynamics of $\phi$ maintains a pdf that is constant
in $\phi$ which matches the assumptions used to derive Eq. \eqref{eq:fokker_planck_rz}.
While the averaged system behaviour remains unchanged, the stochastic
trajectories differ significantly due to measurements, shown in Figure
\ref{fig:trajectories_eigenstates}. Note that each individual trajectory,
if run for a sufficiently long time, would dwell in the vicinity of
either of the boundaries $r_{z}=\pm1$, jumping between the two. However,
once the system reaches stationarity (equilibrium), the ensemble of
trajectories reaches the stationary distribution $p_{st}(r_{z})_{\gamma}$
associated with the corresponding value of $\gamma$. One would expect
this distribution to match the distribution following from solving
the Fokker-Planck equation from Section \ref{sec:fokker_planck_system}.
As can be seen from Figure \ref{fig:trajectories_eigenstates} (right
panel), this is indeed the case in our simulations. 
This demonstrates the stochastic behaviour of the system and the effect
of the measurement interactions that send $r_{z}$ to the vicinity
of the eigenstates, while maintaining the same thermal averaged state
throughout the evolution. This exemplifies how crucial it is to represent
the stochastic behaviour of quantum measurements, and to recognise
that all physical changes in the system are stochastic in nature.
We then show in Figure \ref{fig:trajectory_bloch_sphere} how the
system explores the surface of the Bloch sphere biased by the Hamiltonian
and under the influence of the environment for a couple of individual
trajectories for protocols M and $\overline{{\rm M}}$ (connecting
and disconnecting the measurement device). The trajectory with $\gamma=1$
explores the Bloch sphere fairly uniformly as seen in Figure \ref{fig:fp_equation},
but for $\gamma=2$, the trajectories tend to dwell near the boundaries
at $r_{z}=\pm1$.

\subsection{Entropy production\label{subsec:Entropy-production}}

\begin{figure}[H]
\centering \includegraphics[width=0.6\linewidth]{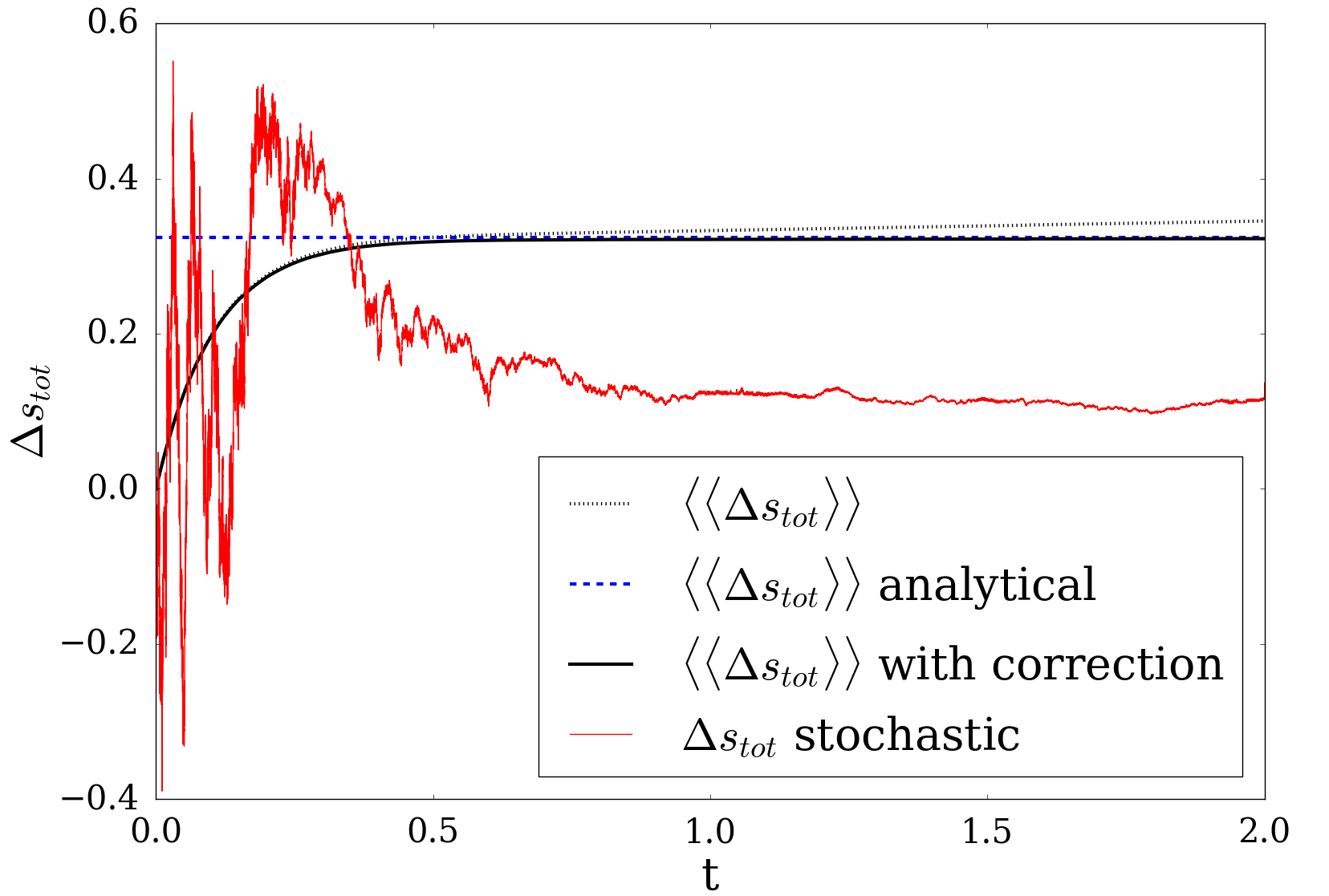}
\caption{Stochastic entropy production for protocol M with $\gamma=2$. The
noise and $r_{z}$ averaged total entropy production is shown with
(solid black) and without (dotted black) the boundary correction.
The red curve corresponds to a single stochastic realisation. The
dashed blue line denotes the asymptotic analytical average total stochastic
entropy production for the process from Eq. \eqref{eq:averaged_total_entropy_production}.
The averages result from one million realisations for $\beta=0.1$,
$\epsilon=1$, $t_{max}=2$, $dt=10^{-5}$, and $\alpha=0.01$.\label{fig:Three-curves-forward}}
\end{figure}

The total stochastic entropy production is obtained by summing up
the stochastic system entropy production in Eq. \eqref{eq:stochastic_system_entropy},
the stochastic environmental entropy production in Eq. \eqref{eq:stochastic_environmental_entropy},
and (for the ensemble average) the boundary correction term from Eq.
\eqref{eq:boundary_term}. The results for protocol M with $\gamma=2$
are shown in Figure \ref{fig:Three-curves-forward}.

As is seen from the Figure, without the boundary correction the mean
entropy production continues to increase which is inconsistent with
a stationary system. The need for boundary corrections is probably
a consequence of our choice of system variables: it arises from the
singular pdf density at the boundaries.

\begin{figure}[H]
\centering\includegraphics[width=0.6\linewidth]{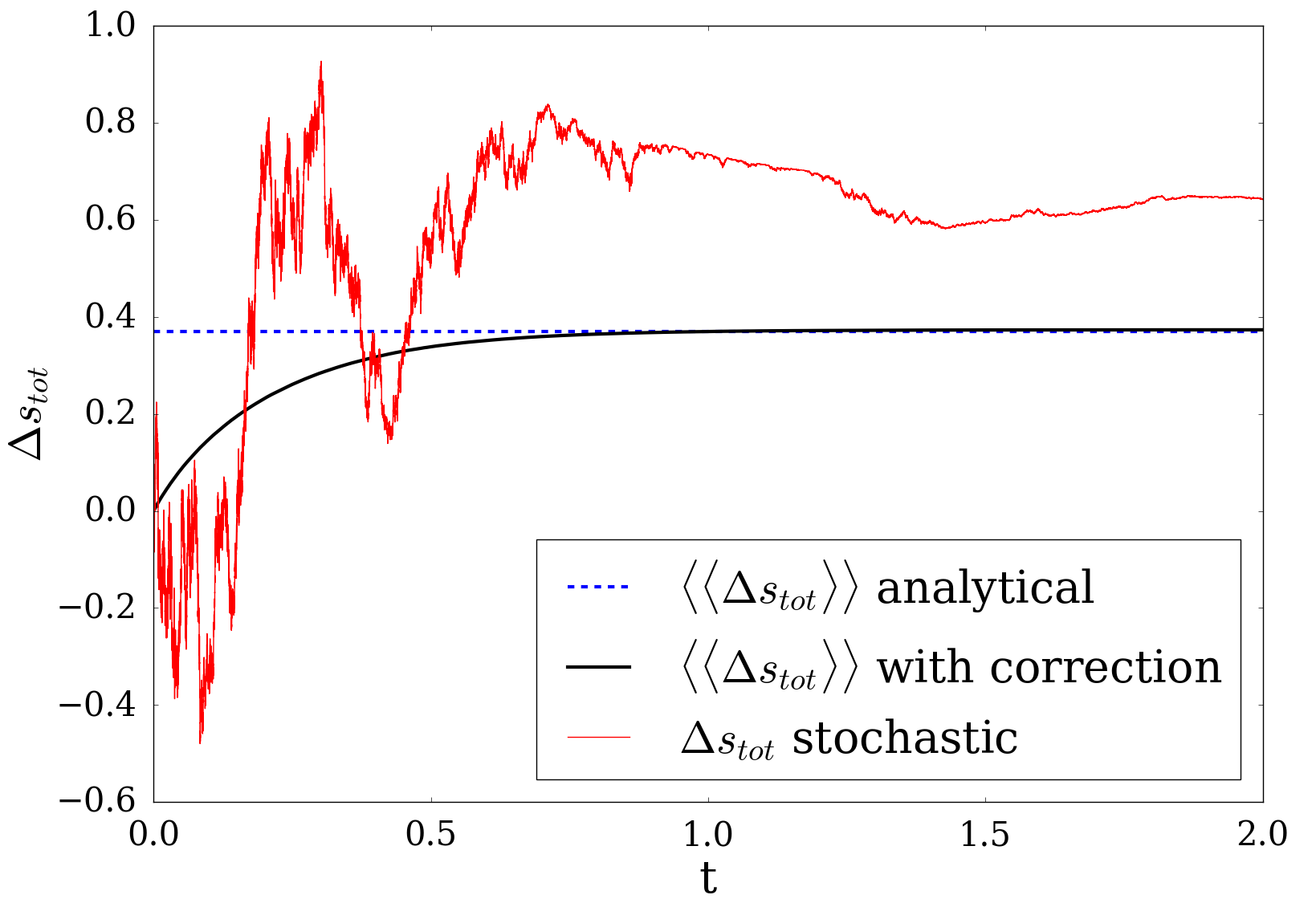}
\caption{The black curve corresponds to noise- and $r_{z}$-averaged total
stochastic entropy production for protocol $\overline{{\rm M}}$,
with $\gamma=1$, including the boundary correction. The red curve
is a single stochastic realisation, and the dashed blue curve is the
asymptotic analytical averaged total stochastic entropy production
from Eq. \eqref{eq:averaged_total_entropy_production}. The averages
are obtained from one million realisations for $\beta=0.1$, $\epsilon=1$,
$t_{max}=2$, $dt=10^{-5}$, and $\alpha=0.01$. The boundary correction
for $\gamma=1$ is significantly smaller than that for $\gamma=2$.\label{fig:two-curves-reverse}}
\end{figure}

The entropy production for protocol $\overline{{\rm M}}$ with $\gamma=1$
is shown in Figure \ref{fig:two-curves-reverse}. As with protocol
M, these results also lead to a ceiling in the average total stochastic
entropy production once the system equilibrates. We see that the process
of disconnecting the measurement device also leads to a positive mean
entropy production; in faimagesct, the mean entropy production is higher
than for the connection process. Note that entropy production tends
to a constant asymptotically even for an individual trajectory. Both
Figs. \ref{fig:Three-curves-forward} and \ref{fig:two-curves-reverse}
display perfect alignment between the analytically averaged total
entropy production for late times, and its numerical counterpart.
This validates the approximations we have made, and the form we have
used for the boundary correction terms.

\subsection{Entropy production pdf and detailed fluctuation theorem}

\begin{figure}[H]
\centering \includegraphics[width=0.6\linewidth]{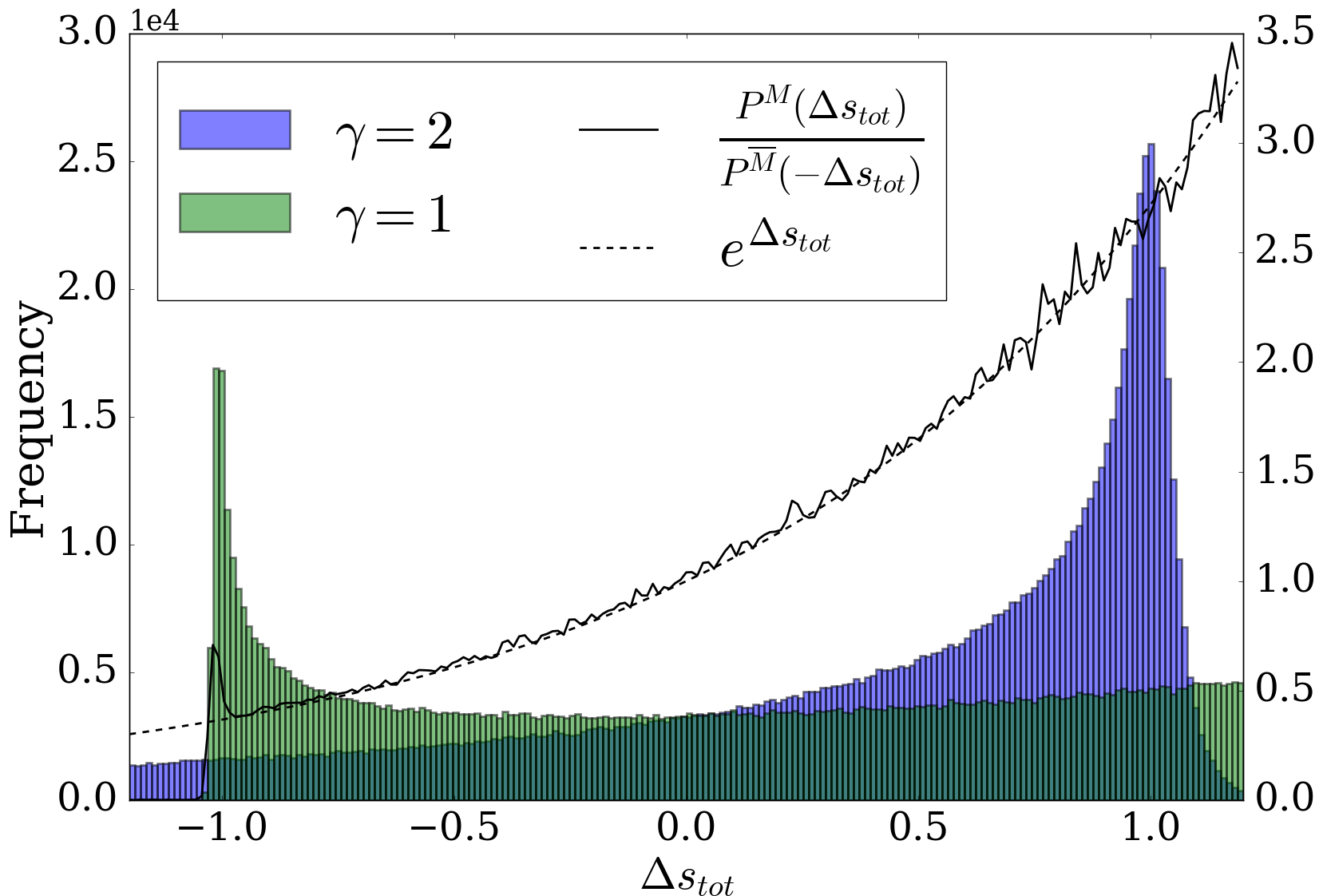}
\caption{Histograms displaying the asymptotic values of the total stochastic
entropy production $\Delta s_{tot}$ for protocol M (blue; $\gamma=2$)
and protocol $\overline{{\rm M}}$ (green; $\gamma=1$) based on one
million individual trajectories (realisations). Black lines show both
sides of Eq. \eqref{eq:fluctuation_relation}, revealing the adherence
to the detailed fluctuation theorem \eqref{eq:fluctuation_relation}.}
\label{fig:fluctuation_relation}
\end{figure}

As a check on the accuracy of the entropy production results, we verify
that the detailed fluctuation theorem Eq. \eqref{eq:fluctuation_relation}
is satisfied. The detailed fluctuation relation concerns processes
described by forward and reverse protocols that are time reversals
of one another, and where the final pdf under the forward protocol
is the same as the initial pdf under the reverse protocol. However,
as we deal here with \emph{distributions} of the entropy production
calculated over \emph{many} trajectories, the strict reversal condition
of the trajectories in protocols M and $\overline{\text{M}}$ is not
necessary and we can compare $e^{\Delta s_{tot}}$ directly with the
ratio of entropy pdfs $P^{M}\left(\Delta s_{tot}\right)/P^{\overline{M}}\left(-\Delta s_{tot}\right)$
. Figure \ref{fig:fluctuation_relation} displays the asymptotic distributions
of total stochastic entropy production for protocols M and $\overline{{\rm M}}$.
These distributions should satisfy the detailed fluctuation relation,
and we indeed find a very good adherence except for deviations at
the extremes of the range due to insufficient sampling.

\section{Discussion and Conclusions\label{sec:Discussion_Conclusion}}

In this paper, we have considered the stochastic entropy production
induced by continuous measurements of a simple open quantum system.
We begin with the stochastic Liouville-von Neumann (SLN) equation
for the system dynamics, in which the environments are represented
by coloured noises, taking the appropriate stochastic average with
respect to these noises and retaining the non-Markovian generality
of the dynamics. Trajectories of the reduced density matrix generated
by the SLN equation are not physical, only the averaged evolution
is of physical significance. In contrast, we require a stochastic
dynamical unravelling that can be interpreted as the outcome of a
physical weak continuous measurement of the system in conjunction
with thermalising influences of the environment. To do so, we take
the Markovian limit of the behaviour of the environment in the SLN
by assuming a high temperature with a specific choice of a two-level
bosonic Hamiltonian $H_{sys}=\epsilon\sigma_{z}$. These steps enable
us to perform a Markovian stochastic unravelling described by an Itô
process, which then allows us to derive an appropriate Fokker-Planck
equation and understand how the pdf of the reduced density matrix
changes with the strength of the continuous measurements while coupled
to a thermalising bath. We have set up a system that interacts with
three independent harmonic baths through the three Pauli matrices,
allowing for a quasi-isotropic stochastic exploration of the Bloch
sphere consistent with residence in a thermal state. We take the view
that a system under constant interaction with an underspecified environment
is best represented by an ensemble of reduced density matrices that
conveys its uncertainty.

The measurement of system energy is then realised by increasing $\gamma$,
the strength of environmental coupling to $\sigma_{z}$, causing the
system to dwell in the vicinity of one of the two energy eigenstates.
We can then obtain the stochastic entropy production associated with
each stochastic trajectory of the system. Note that if the pdf of
the reduced density matrix is stationary for a particular set of environmental
couplings, raising $\gamma$ will cause the pdf to increase in the
vicinity of the eigenstates of the system Hamiltonian but this will
not change the \emph{average} energy of an ensemble of systems. A
model of the dynamics and thermodynamics of measurement requires the
system to be represented by a \emph{member} of an ensemble: the average
behaviour will not suffice.

To calculate the entropy production associated with measurement, we
use an analysis of the Markovian system dynamics developed in \cite{spinney2012entropy}.
This contrasts with the calculation of the stochastic entropy production
for quantum systems based on forward and reverse trajectories constructed
using forward and reverse Kraus operators \cite{elouard2017role,manikandan2019fluctuation,crooks2008quantum}.
We also compute system entropy production by considering the evolution
of the pdf of the reduced density matrix of the system.

We find that the numerically calculated entropy production contains
some subtle numerical artefacts. These are corrected by calculating
the mean analytical system entropy production in a stationary state,
allowing us to obtain boundary correction terms that eliminate the
artefacts.

We find that the increase and decrease in coupling strength $\gamma$,
corresponding to the operation of a measuring device, are both accompanied
by a positive mean stochastic entropy production. This is the entropic
cost of quantum measurement. Furthermore, we show that the stochastic
entropy production associated with a quantum measurement is finite
as the system achieves a new stationary state. Finally, we have shown
that the processes of attachment and detachment of a measuring device
generate distributions of stochastic entropy production that satisfy
a detailed fluctuation relation. We aim to extend this approach next
to more complicated systems, to include non-Markovian dynamics and
to relax the requirement for a high temperature approximation.

\section*{Acknowledgments}

D. Matos is supported by the EPSRC Centre for Doctoral Training in
Cross-Disciplinary Approaches to Non-Equilibrium Systems (CANES, Grant
No. EP/L015854/1). Calculations in this paper were performed using
the King’s College HPC cluster Gravity.

\appendix

\section{Appendices}

\subsection{Averaged SLN equation\label{subsec:Averaged-SLN-equation}}

Here we shall consider the calculation of the functional derivatives
of $\rho_{s}$ with respect to the noises needed for the derivation
of Eq. \eqref{eq:averaged_sln_general}. The formal solution of the
SLN equation is given by 
\begin{equation}
\rho_{s}(t)=U_{+}(t,0)\rho_{s}(0)U_{-}(0,t),\label{eq:}
\end{equation}
where $U_{+}$ and $U_{-}$ are the appropriate forward and backward
propagators defined as

\begin{eqnarray}
U_{+}(t,0) & = & \widehat{\mathcal{T}}_{+}\exp\left[-i\int_{0}^{t}dt^{\prime}H_{+}(t^{\prime})\right],\\
U_{-}(0,t) & = & \widehat{\mathcal{T}}_{-}\exp\left[i\int_{0}^{t}dt^{\prime}H_{-}(t^{\prime})\right],
\end{eqnarray}
where $\widehat{\mathcal{T}}_{+}$ and $\widehat{\mathcal{T}}_{-}$
are the forward and backward time ordering operators, respectively,
and
\begin{equation}
H_{\pm}(t)=H_{sys}(t)-\sum_{k}\left(\eta_{k}(t)\pm\frac{\nu_{k}(t)}{2}\right)f_{k}\label{eq:-1}
\end{equation}
are the Hamiltonian operators. Note that they are not adjoints of
each other, see Ref. \cite{mccaul2017partition} for details. Hence
the functional derivative of the density matrix can be written as

\begin{equation}
\frac{\delta\rho_{s}(t)}{\delta\zeta(t^{\prime})}=\frac{\delta U_{+}(t,0)}{\delta\zeta(t^{\prime})}\rho_{s}(0)U_{-}(0,t)+U_{+}(t,0)\rho_{s}(0)\frac{\delta U_{-}(0,t)}{\delta\zeta(t^{\prime})}.
\end{equation}
The derivatives of the forward and backward propagators with respect
to arbitrary perturbations of the noises are given by:

\begin{align}
\frac{\delta U_{+}(t,t^{\prime})}{\delta\zeta(\tau^{\prime})} & =-i\int_{t^{\prime}}^{t}U_{+}(t,\tau)\frac{\delta V_{+}(\tau)}{\delta\zeta(\tau^{\prime})}U_{+}(\tau,t^{\prime})d\tau\\
\frac{\delta U_{-}(t^{\prime},t)}{\delta\zeta(\tau^{\prime})} & =i\int_{t^{\prime}}^{t}U_{-}(t^{\prime},\tau)\frac{\delta V_{-}(\tau)}{\delta\zeta(\tau^{\prime})}U_{-}(\tau,t)d\tau,
\end{align}
where $V_{\pm}(\tau)=-\left(\eta(\tau)\pm\frac{\nu(\tau)}{2}\right)f$,
for convenience dropping the suffix $k.$ We can now define the variation
of the propagators with respect to the noises $\eta(t)$ and $\nu(t)$.
Starting with $\eta(t)$:

\begin{align}
\frac{\delta U_{+}(t,t^{\prime})}{\delta\eta(\tau)} & =-i\int_{t^{\prime}}^{t}U_{+}(t,\tau^{\prime})\left[-\delta(\tau-\tau^{\prime})f(\tau^{\prime})\right]U_{+}(\tau^{\prime},t^{\prime})d\tau^{\prime}\\
 & =iU_{+}(t,\tau)f(\tau)U_{+}(\tau,t^{\prime})\,,\nonumber \\
\frac{\delta U_{-}(t^{\prime},t)}{\delta\eta(\tau)} & =i\int_{t^{\prime}}^{t}U_{-}(t^{\prime},\tau^{\prime})\left[-\delta(\tau-\tau^{\prime})f(\tau^{\prime})\right]U_{-}(\tau^{\prime},t)d\tau^{\prime}\nonumber \\
 & =-iU_{-}(t^{\prime},\tau)f(\tau)U_{-}(\tau,t).\nonumber 
\end{align}
For $\nu(t)$:

\begin{align}
\frac{\delta U_{+}(t,t^{\prime})}{\delta\nu(\tau)} & =-i\int_{t^{\prime}}^{t}U_{+}(t,\tau^{\prime})\left[-\frac{1}{2}\delta(\tau-\tau^{\prime})f(\tau^{\prime})\right]U_{+}(\tau^{\prime},t^{\prime})d\tau^{\prime}\\
 & =\frac{i}{2}U_{+}(t,\tau)f(\tau)U_{+}(\tau,t^{\prime})\,,\nonumber \\
\frac{\delta U_{-}(t^{\prime},t)}{\delta\nu(\tau)} & =i\int_{t^{\prime}}^{t}U_{-}(t^{\prime},\tau^{\prime})\left[\frac{1}{2}\delta(\tau-\tau^{\prime})f(\tau^{\prime})\right]U_{-}(\tau^{\prime},t)d\tau^{\prime}\nonumber \\
 & =\frac{i}{2}U_{-}(t^{\prime},\tau)f(\tau)U_{-}(\tau,t).\nonumber 
\end{align}
This then allows us to write $\left\langle \eta(t)\rho_{s}(t)\right\rangle $
as

\begin{align}
\langle\eta(t)\rho_{s}(t)\rangle & =\int_{0}^{t}dt^{\prime}K^{\eta\eta}(t-t^{\prime})\left\langle \frac{\delta\rho_{s}(t)}{\delta\eta(t^{\prime})}\right\rangle +\int_{0}^{t}dt^{\prime}K^{\eta\nu}(t-t^{\prime})\left\langle \frac{\delta\rho_{s}(t)}{\delta\nu(t^{\prime})}\right\rangle \nonumber \\
 & =\int_{0}^{t}dt^{\prime}K^{\eta\eta}(t-t^{\prime})\left\langle iU_{+}(t,t^{\prime})f(t^{\prime})U_{+}(t^{\prime},0)\rho_{s}(0)U_{-}(0,t)-U_{+}(t,0)\rho_{s}(0)iU_{-}(0,t^{\prime})f(t^{\prime})U_{-}(t^{\prime},t)\right\rangle \\
 & \qquad+\int_{0}^{t}dt^{\prime}K^{\eta\nu}(t-t^{\prime})\left\langle \frac{i}{2}U_{+}(t,t^{\prime})f(t^{\prime})U_{+}(t^{\prime},0)\rho_{s}(0)U_{-}(0,t)+U_{+}(t,0)\rho_{s}(0)\frac{i}{2}U_{-}(0,t^{\prime})f(t^{\prime})U_{-}(t^{\prime},t)\right\rangle .\nonumber 
\end{align}
Using this expression, we arrive at Eq. \eqref{eq:averaged_sln_general}
given in the text.

\subsection{Cylindrical coordinate equations of motion\label{subsec:Cylindrical-coordinate-equations}}

In this section, we derive the equations of motion in cylindrical
coordinates $(r^{2},r_{z},\phi)$ for general $\gamma$. With these
coordinates, the $x$ and $y$ components are given by

\begin{equation}
r_{x}=\sqrt{r^{2}-r_{z}^{2}}\cos\phi\hspace{5em}r_{y}=\sqrt{r^{2}-r_{z}^{2}}\sin\phi.\label{eq:cylindrical_coords-1}
\end{equation}
From these expressions and using Itô's Lemma, we can obtain the equations
of motion for this set of coordinates:

\begin{align}
dr^{2} & =\frac{\partial r^{2}}{\partial r_{z}}dr_{z}+\frac{1}{2}\frac{\partial^{2}r^{2}}{\partial r_{z}^{2}}\left(dr_{z}\right)^{2}+\frac{\partial r^{2}}{\partial r_{y}}dr_{y}+\frac{1}{2}\frac{\partial r^{2}}{\partial r_{y}^{2}}\left(dr_{y}\right)^{2}+\frac{\partial r^{2}}{\partial r_{x}}dr_{x}+\frac{1}{2}\frac{\partial^{2}r^{2}}{\partial r_{x}^{2}}\left(dr_{x}\right)^{2}\label{eq:dr_squared_ito}\\
d\phi & =\frac{\partial\phi}{\partial r_{x}}dr_{x}+\frac{1}{2}\frac{\partial^{2}\phi}{\partial r_{x}^{2}}\left(dr_{x}\right)^{2}+\frac{\partial\phi}{\partial r_{y}}dr_{y}+\frac{1}{2}\frac{\partial^{2}\phi}{\partial r_{y}^{2}}\left(dr_{y}\right)^{2}+\frac{\partial^{2}\phi}{\partial r_{x}\partial r_{y}}dr_{x}dr_{y}.\label{eq:dphi_ito}
\end{align}
The required derivatives are:

\begin{gather}
\frac{\partial r^{2}}{\partial r_{x}}=2r_{x}\qquad\frac{\partial r^{2}}{\partial r_{y}}=2r_{y}\qquad\frac{\partial r^{2}}{\partial r_{z}}=2r_{z}\\
\frac{\partial^{2}r^{2}}{\partial r_{x}^{2}}=\frac{\partial^{2}r^{2}}{\partial r_{y}^{2}}=\frac{\partial^{2}r^{2}}{\partial r_{z}^{2}}=2\\
\frac{\partial\phi}{\partial r_{x}}=-\frac{r_{y}}{r_{x}^{2}+r_{y}^{2}}\qquad\frac{\partial\phi}{\partial r_{y}}=\frac{r_{x}}{r_{x}^{2}+r_{y}^{2}}\\
\frac{\partial^{2}\phi}{\partial r_{x}^{2}}=\frac{2r_{x}r_{y}}{\left(r_{x}^{2}+r_{y}^{2}\right)^{2}}\qquad\frac{\partial^{2}\phi}{\partial r_{y}^{2}}=-\frac{2r_{x}r_{y}}{\left(r_{x}^{2}+r_{y}^{2}\right)^{2}}\qquad\frac{\partial^{2}\phi}{\partial r_{x}r_{y}}=\frac{r_{y}^{2}-r_{x}^{2}}{\left(r_{x}^{2}+r_{y}^{2}\right)^{2}},
\end{gather}
which allows us to simplify Eqs. \eqref{eq:dr_squared_ito} and \eqref{eq:dphi_ito}:

\begin{align}
dr^{2} & =2r_{z}dr_{z}+\left(dr_{z}\right)^{2}+2r_{y}dr_{y}+\left(dr_{y}\right)^{2}+2r_{x}dr_{x}+\left(dr_{x}\right)^{2}\\
d\phi & =-\frac{r_{y}}{r_{x}^{2}+r_{y}^{2}}dr_{x}+\frac{r_{x}r_{y}}{\left(r_{x}^{2}+r_{y}^{2}\right)^{2}}\left(dr_{x}\right)^{2}+\frac{r_{x}}{r_{x}^{2}+r_{y}^{2}}dr_{y}-\frac{r_{x}r_{y}}{\left(r_{x}^{2}+r_{y}^{2}\right)^{2}}\left(dr_{y}\right)^{2}+\frac{r_{y}^{2}-r_{x}^{2}}{\left(r_{x}^{2}+r_{y}^{2}\right)^{2}}dr_{x}dr_{y}.
\end{align}
We can calculate $(dr_{x})^{2}$, $(dr_{y})^{2}$, $(dr_{z})^{2}$
and $dr_{x}dr_{y}$ using Eqs. \eqref{eq:dr_x_general}-\eqref{eq:dr_z_general}
and the substitution $dW_{i}dW_{j}=\delta_{ij}dt$, keeping linear
terms in $dt$:

\begin{align}
(dr_{x})^{2} & =\lambda^{2}\left(\beta\epsilon r_{z}+2(1-r_{x}^{2})\right)^{2}dt+4\lambda^{2}r_{x}^{2}r_{y}^{2}dt+4\lambda^{2}\gamma^{2}r_{x}^{2}r_{z}^{2}dt\\
(dr_{y})^{2} & =4\lambda^{2}r_{x}^{2}r_{y}^{2}dt+\lambda^{2}\left(\beta\epsilon r_{z}+2(1-r_{y}^{2})\right)^{2}dt+4\lambda^{2}\gamma^{2}r_{y}^{2}r_{z}^{2}dt\\
(dr_{z})^{2} & =\lambda^{2}r_{x}^{2}\left(\beta\epsilon+2r_{z}\right)^{2}dt+\lambda^{2}r_{y}^{2}\left(\beta\epsilon+2r_{z}\right)^{2}dt+4\lambda^{2}\gamma^{2}\left(1-r_{z}^{2}\right)^{2}dt\\
dr_{x}dr_{y} & =-2\lambda^{2}r_{x}r_{y}\left(\beta\epsilon r_{z}+2(1-r_{x}^{2})\right)dt-2\lambda^{2}r_{x}r_{y}\left(\beta\epsilon r_{z}+2(1-r_{y}^{2})\right)dt+4\lambda^{2}\gamma^{2}r_{x}r_{y}r_{z}^{2}dt.
\end{align}
After some algebra and dropping quadratic terms in $\beta$, we obtain

\begin{align}
(dr_{x})^{2} & =4\lambda^{2}\left[r_{x}^{2}\left(r_{x}^{2}+r_{y}^{2}+\gamma^{2}r_{z}^{2}\right)+\beta\epsilon r_{z}\left(1-r_{x}^{2}\right)+1-2r_{x}^{2}\right]dt\nonumber \\
(dr_{y})^{2} & =4\lambda^{2}\left[r_{y}^{2}\left(r_{x}^{2}+r_{y}^{2}+\gamma^{2}r_{z}^{2}\right)+\beta\epsilon r_{z}\left(1-r_{y}^{2}\right)+1-2r_{y}^{2}\right]dt\label{eq:dri_squared_ito_eqs}\\
(dr_{z})^{2} & =4\lambda^{2}\left[\beta\epsilon r_{z}\left(r_{x}^{2}+r_{y}^{2}\right)+r_{z}^{2}\left(r_{x}^{2}+r_{y}^{2}\right)-2\gamma^{2}r_{z}^{2}+\gamma^{2}\left(1+r_{z}^{4}\right)\right]dt.\nonumber 
\end{align}
The sum of Eqs. \eqref{eq:dri_squared_ito_eqs} is given by

\begin{align}
(dr_{x})^{2}+(dr_{y})^{2}+(dr_{z})^{2} & =4\lambda^{2}\left[-2\gamma^{2}r_{z}^{2}+2\beta\epsilon r_{z}+\gamma^{2}\left(1+r_{z}^{4}\right)\right.\label{eq:sum_drs_ito}\\
 & \left.\qquad\qquad\qquad+\left(r^{2}-r_{z}^{2}\right)\left(r^{2}+\gamma^{2}r_{z}^{2}-2\right)+2\right]dt.\nonumber 
\end{align}
Moving on to the $2r_{i}dr_{i}$ terms:

\begin{align}
2r_{x}dr_{x} & =-4\epsilon r_{x}r_{y}dt-4\lambda^{2}\left(1+\gamma^{2}\right)r_{x}^{2}dt+2\lambda r_{x}\left(\beta\epsilon r_{z}+2\left(1-r_{x}^{2}\right)\right)dW_{x}-4\lambda r_{x}^{2}r_{y}dW_{y}-4\gamma\lambda r_{x}^{2}r_{z}dW_{z}\nonumber \\
2r_{y}dr_{y} & =4\epsilon r_{x}r_{y}dt-4\lambda^{2}\left(1+\gamma^{2}\right)r_{y}^{2}dt-4\lambda r_{x}r_{y}^{2}dW_{x}+2\lambda r_{y}\left(\beta\epsilon r_{z}+2\left(1-r_{y}^{2}\right)\right)dW_{y}-4\gamma\lambda r_{y}^{2}r_{z}dW_{z}\label{eq:2rdr_ito_eqs}\\
2r_{z}dr_{z} & =-8\lambda^{2}r_{z}^{2}dt-8\lambda^{2}\beta\epsilon r_{z}dt-2\lambda r_{x}r_{z}\left(\beta\epsilon+2r_{z}\right)dW_{x}-2\lambda r_{y}r_{z}\left(\beta\epsilon+2r_{z}\right)dW_{y}+4\gamma\lambda r_{z}\left(1-r_{z}^{2}\right)dW_{z}.\nonumber 
\end{align}
Adding up Eqs. \eqref{eq:2rdr_ito_eqs} leads to
\begin{equation}
\begin{aligned}2r_{x}dr_{x}+2r_{y}dr_{y}+2r_{z}dr_{z} & =-4\lambda^{2}\left(2r_{z}^{2}+(1+\gamma^{2})(r^{2}-r_{z}^{2})\right)dt-8\lambda^{2}\beta\epsilon r_{z}dt+4\lambda r_{x}\left(1-r^{2}\right)dW_{x}\\
 & \qquad+4\lambda r_{y}\left(1-r^{2}\right)dW_{y}+4\lambda\gamma r_{z}\left(1-r^{2}\right)dW_{z}.
\end{aligned}
\label{eq:sums_rdrs_ito}
\end{equation}

By adding Eqs. \eqref{eq:sum_drs_ito} and \eqref{eq:sums_rdrs_ito}
we obtain the final expression for $dr^{2}$:

\begin{align}
\begin{aligned}dr^{2} & =4\lambda^{2}\left(r^{2}-1\right)\left(\gamma^{2}r_{z}^{2}-\gamma^{2}+r^{2}-r_{z}^{2}-2\right)dt+4\lambda\sqrt{1-\frac{r_{z}^{2}}{r^{2}}}r\cos\phi\left(1-r^{2}\right)dW_{x}\\
 & +4\lambda\sqrt{1-\frac{r_{z}^{2}}{r^{2}}}r\sin\phi\left(1-r^{2}\right)dW_{y}+4\lambda\gamma r_{z}\left(1-r^{2}\right)dW_{z}\,.
\end{aligned}
\end{align}
Moving on to calculating $d\phi$, the contributions linear in $dr_{i}$
are:

\begin{equation}
-\frac{r_{y}}{r_{x}^{2}+r_{y}^{2}}dr_{x}+\frac{r_{x}}{r_{x}^{2}+r_{y}^{2}}dr_{y}=\frac{1}{r_{x}^{2}+r_{y}^{2}}\left[2\epsilon\left(r_{x}^{2}+r_{y}^{2}\right)dt-\lambda r_{y}\left(\beta\epsilon r_{z}+2\right)dW_{x}+\lambda r_{x}\left(\beta\epsilon r_{z}+2\right)dW_{y}\right].\label{eq:dr2}
\end{equation}
The terms quadratic in $dr_{i}$ lead to

\begin{equation}
\frac{r_{x}r_{y}}{\left(r_{x}^{2}+r_{y}^{2}\right)^{2}}\left(dr_{x}\right)^{2}-\frac{r_{x}r_{y}}{\left(r_{x}^{2}+r_{y}^{2}\right)^{2}}\left(dr_{y}\right)^{2}+\frac{r_{y}^{2}-r_{x}^{2}}{\left(r_{x}^{2}+r_{y}^{2}\right)^{2}}dr_{x}dr_{y}=0,\label{eq:dri}
\end{equation}
allowing us to write the final expression for $d\phi$ as

\begin{equation}
d\phi=2\epsilon dt-\lambda\frac{\left(\beta\epsilon r_{z}+2\right)\sin\phi}{\sqrt{r^{2}-r_{z}^{2}}}dW_{x}+\lambda\frac{\left(\beta\epsilon r_{z}+2\right)\cos\phi}{\sqrt{r^{2}-r_{z}^{2}}}dW_{y}.\label{eq:dphi}
\end{equation}

\subsection{Averaged system entropy production and boundary terms\label{subsec:Averaged-system-entropy}}

We derive an analytical expression for $d\langle\langle\Delta s_{sys}\rangle\rangle$,
the noise- and coordinate-averaged, incremental system entropy production,
written as

\begin{align}
d\langle\langle\Delta s_{sys}\rangle\rangle & =\int_{-1}^{1}d\langle\Delta s_{sys}\rangle\,p(r_{z},t)dr_{z},
\end{align}
where $d\langle\Delta s_{sys}\rangle$ is given by the noise averaged
form of Eq. \eqref{eq:stochastic_system_entropy}. We proceed as follows:
\begin{align}
d\langle\langle\Delta s_{sys}\rangle\rangle & =\int_{-1}^{1}\left(-\frac{\partial\ln p(r_{z},t)}{\partial t}dt-\frac{\partial\ln p(r_{z},t)}{\partial r_{z}}d\langle r_{z}\rangle-D_{zz}\frac{\partial^{2}\ln p(r_{z},t)}{\partial r_{z}^{2}}dt\right)p(r_{z},t)dr_{z}\nonumber \\
 & =\int_{-1}^{1}\left(-\frac{\partial\ln p(r_{z},t)}{\partial t}dt-A_{z}\frac{\partial\ln p(r_{z},t)}{\partial r_{z}}dt-D_{zz}\frac{\partial^{2}\ln p(r_{z},t)}{\partial r_{z}^{2}}dt\right)p(r_{z},t)dr_{z}.\label{eq:a3.2}
\end{align}
Integrating the last term by parts, and introducing the $r_{z}$ component
of the probability current, 
\begin{equation}
J_{z}=A_{z}p(r_{z},t)-\frac{\partial\left(D_{zz}p(r_{z},t)\right)}{\partial r_{z}}\,,\label{eq:a3.3}
\end{equation}
that appears in the Fokker-Planck equation \eqref{eq:fokker_planck_eq_multidim}
as 
\begin{equation}
\frac{\partial p(r_{z},t)}{\partial t}=-\frac{\partial J_{z}}{\partial r_{z}}\,,\label{eq:FP}
\end{equation}
we can write:

\begin{align}
d\langle\langle\Delta s_{sys}\rangle\rangle & =\int_{-1}^{1}\left(\left(-\frac{\partial\ln p(r_{z},t)}{\partial t}dt-A_{z}\frac{\partial\ln p(r_{z},t)}{\partial r_{z}}dt\right)p(r_{z},t)+\frac{\partial\left(D_{zz}p(r_{z},t)\right)}{\partial r_{z}}\frac{\partial\ln p(r_{z},t)}{\partial r_{z}}dt\right)dr_{z}\nonumber \\
 & \qquad\qquad-\left[D_{zz}\frac{\partial\ln p(r_{z},t)}{\partial r_{z}}p(r_{z},t)\right]_{-1}^{1}dt\label{eq:a3.4}\\
 & =\int_{-1}^{1}\left(-\frac{\partial\ln p(r_{z},t)}{\partial t}p(r_{z},t)dt-J_{z}\frac{\partial\ln p(r_{z},t)}{\partial r_{z}}dt\right)dr_{z}-\left[D_{zz}\frac{\partial p(r_{z},t)}{\partial r_{z}}\right]_{-1}^{1}dt.\nonumber 
\end{align}
Next we integrate the second term inside the integral by parts and
replace the $r_{z}$ derivative of the probability current with the
left hand side of the Fokker-Planck equation \eqref{eq:FP}:
\begin{align}
d\langle\langle\Delta s_{sys}\rangle\rangle & =\int_{-1}^{1}\left(-\frac{\partial\ln p(r_{z},t)}{\partial t}p(r_{z},t)dt+\ln p(r_{z},t)\,\frac{\partial J_{z}}{\partial r_{z}}dt\right)dr_{z}-\left[J_{z}\ln p(r_{z},t)\right]_{-1}^{1}dt-\left[D_{zz}\frac{\partial p(r_{z},t)}{\partial r_{z}}\right]_{-1}^{1}dt\nonumber \\
 & =-\int_{-1}^{1}\left(\frac{\partial\ln p(r_{z},t)}{\partial t}p(r_{z},t)dt+\ln p(r_{z},t)\,\frac{\partial p(r_{z},t)}{\partial t}dt\right)dr_{z}-\left[J_{z}\ln p(r_{z},t)\right]_{-1}^{1}dt-\left[D_{zz}\frac{\partial p(r_{z},t)}{\partial r_{z}}\right]_{-1}^{1}dt\nonumber \\
 & =-\left(\frac{d}{dt}\int_{-1}^{1}p(r_{z},t)\ln p(r_{z},t)dr_{z}\right)dt-\left[J_{z}\ln p(r_{z},t)\right]_{-1}^{1}dt-\left[D_{zz}\frac{\partial p(r_{z},t)}{\partial r_{z}}\right]_{-1}^{1}dt\nonumber \\
 & =dS_{G}(t)-\left[J_{z}\ln p(r_{z},t)\right]_{-1}^{1}dt-\left[D_{zz}\frac{\partial p(r_{z},t)}{\partial r_{z}}\right]_{-1}^{1}dt,\label{eq:a3.5}
\end{align}
where $dS_{G}$ is the increment of the Gibbs entropy, see Eq. \eqref{eq:averaged_system_entropy_production}.

\bibliographystyle{unsrt}
\bibliography{paper_entropy_production_20220515}

\begin{thebibliography}{10}

\bibitem{shor1995scheme}
Peter~W Shor.
\newblock Scheme for reducing decoherence in quantum computer memory.
\newblock {\em Physical review A}, 52(4):R2493, 1995.

\bibitem{weiss2012quantum}
Ulrich Weiss.
\newblock {\em Quantum dissipative systems}, volume~13.
\newblock World scientific, 2012.

\bibitem{sekimoto1998langevin}
Ken Sekimoto.
\newblock Langevin equation and thermodynamics.
\newblock {\em Progress of Theoretical Physics Supplement}, 130:17--27, 1998.

\bibitem{seifert2005entropy}
Udo Seifert.
\newblock Entropy production along a stochastic trajectory and an integral
  fluctuation theorem.
\newblock {\em Physical review letters}, 95(4):040602, 2005.

\bibitem{spinney2012entropy}
Richard~E Spinney and Ian~J Ford.
\newblock Entropy production in full phase space for continuous stochastic
  dynamics.
\newblock {\em Physical Review E}, 85(5):051113, 2012.

\bibitem{carmichael2009open}
Howard Carmichael.
\newblock {\em An open systems approach to quantum optics: lectures presented
  at the Universit{\'e} Libre de Bruxelles, October 28 to November 4, 1991},
  volume~18.
\newblock Springer Science \& Business Media, 2009.

\bibitem{breuer2002theory}
Heinz-Peter Breuer, Francesco Petruccione, et~al.
\newblock {\em The theory of open quantum systems}.
\newblock Oxford University Press on Demand, 2002.

\bibitem{stockburger2001non}
J{\"u}rgen~T Stockburger and Hermann Grabert.
\newblock Non-markovian quantum state diffusion.
\newblock {\em Chemical Physics}, 268(1-3):249--256, 2001.

\bibitem{stockburger2002exact}
J{\"u}rgen~T Stockburger and Hermann Grabert.
\newblock Exact c-number representation of non-markovian quantum dissipation.
\newblock {\em Physical review letters}, 88(17):170407, 2002.

\bibitem{stockburger2004simulating}
J{\"u}rgen~T Stockburger.
\newblock Simulating spin-boson dynamics with stochastic liouville--von neumann
  equations.
\newblock {\em Chemical physics}, 296(2-3):159--169, 2004.

\bibitem{mccaul2017partition}
GMG McCaul, CD~Lorenz, and L~Kantorovich.
\newblock Partition-free approach to open quantum systems in harmonic
  environments: An exact stochastic liouville equation.
\newblock {\em Physical Review B}, 95(12):125124, 2017.

\bibitem{lane2020exactly}
MA~Lane, D~Matos, IJ~Ford, and L~Kantorovich.
\newblock Exactly thermalised quantum dynamics of the spin-boson model coupled
  to a dissipative environment.
\newblock {\em arXiv preprint arXiv:2002.07700}, 2020.

\bibitem{feynman1963theory}
Richard~Phillips Feynman and FL~Vernon~Jr.
\newblock The theory of a general quantum system interacting with a linear
  dissipative system.
\newblock {\em Annals of physics}, 24:118--173, 1963.

\bibitem{evans1993probability}
Denis~J Evans, Ezechiel Godert~David Cohen, and Gary~P Morriss.
\newblock Probability of second law violations in shearing steady states.
\newblock {\em Physical review letters}, 71(15):2401, 1993.

\bibitem{evans1995steady}
Denis~J Evans and Debra~J Searles.
\newblock Steady states, invariant measures, and response theory.
\newblock {\em Physical Review E}, 52(6):5839, 1995.

\bibitem{evans2002fluctuation}
Denis~J Evans and Debra~J Searles.
\newblock The fluctuation theorem.
\newblock {\em Advances in Physics}, 51(7):1529--1585, 2002.

\bibitem{carberry2004fluctuations}
DM~Carberry, James~Cowie Reid, GM~Wang, Edith~M Sevick, Debra~J Searles, and
  Denis~J Evans.
\newblock Fluctuations and irreversibility: An experimental demonstration of a
  second-law-like theorem using a colloidal particle held in an optical trap.
\newblock {\em Physical review letters}, 92(14):140601, 2004.

\bibitem{gallavotti1995dynamical}
Giovanni Gallavotti and Ezechiel Godert~David Cohen.
\newblock Dynamical ensembles in nonequilibrium statistical mechanics.
\newblock {\em Physical review letters}, 74(14):2694, 1995.

\bibitem{kurchan1998fluctuation}
Jorge Kurchan.
\newblock Fluctuation theorem for stochastic dynamics.
\newblock {\em Journal of Physics A: Mathematical and General}, 31(16):3719,
  1998.

\bibitem{lebowitz1999gallavotti}
Joel~L Lebowitz and Herbert Spohn.
\newblock A gallavotti--cohen-type symmetry in the large deviation functional
  for stochastic dynamics.
\newblock {\em Journal of Statistical Physics}, 95(1):333--365, 1999.

\bibitem{jarzynski1997nonequilibrium}
Christopher Jarzynski.
\newblock Nonequilibrium equality for free energy differences.
\newblock {\em Physical Review Letters}, 78(14):2690, 1997.

\bibitem{crooks1998nonequilibrium}
Gavin~E Crooks.
\newblock Nonequilibrium measurements of free energy differences for
  microscopically reversible markovian systems.
\newblock {\em Journal of Statistical Physics}, 90(5):1481--1487, 1998.

\bibitem{crooks1999entropy}
Gavin~E Crooks.
\newblock Entropy production fluctuation theorem and the nonequilibrium work
  relation for free energy differences.
\newblock {\em Physical Review E}, 60(3):2721, 1999.

\bibitem{harris2007fluctuation}
Rosemary~J Harris and Gunther~M Sch{\"u}tz.
\newblock Fluctuation theorems for stochastic dynamics.
\newblock {\em Journal of Statistical Mechanics: Theory and Experiment},
  2007(07):P07020, 2007.

\bibitem{ford2015stochastic}
Ian~J Ford, Zachary~PL Laker, and Henry~J Charlesworth.
\newblock Stochastic entropy production arising from nonstationary thermal
  transport.
\newblock {\em Physical review E}, 92(4):042108, 2015.

\bibitem{jacobs2014quantum}
Kurt Jacobs.
\newblock {\em Quantum measurement theory and its applications}.
\newblock Cambridge University Press, 2014.

\bibitem{jacobs2006straightforward}
Kurt Jacobs and Daniel~A Steck.
\newblock A straightforward introduction to continuous quantum measurement.
\newblock {\em Contemporary Physics}, 47(5):279--303, 2006.

\bibitem{crooks2008quantum}
Gavin~E Crooks.
\newblock Quantum operation time reversal.
\newblock {\em Physical Review A}, 77(3):034101, 2008.

\bibitem{horowitz2013entropy}
Jordan~M Horowitz and Juan~MR Parrondo.
\newblock Entropy production along nonequilibrium quantum jump trajectories.
\newblock {\em New Journal of Physics}, 15(8):085028, 2013.

\bibitem{leggio2013entropy}
B~Leggio, A~Napoli, A~Messina, and H-P Breuer.
\newblock Entropy production and information fluctuations along quantum
  trajectories.
\newblock {\em Physical Review A}, 88(4):042111, 2013.

\bibitem{elouard2017role}
Cyril Elouard, David~A Herrera-Mart{\'\i}, Maxime Clusel, and Alexia Auffeves.
\newblock The role of quantum measurement in stochastic thermodynamics.
\newblock {\em npj Quantum Information}, 3(1):1--10, 2017.

\bibitem{elouard2017probing}
Cyril Elouard, NK~Bernardes, ARR Carvalho, MF~Santos, and A~Auff{\`e}ves.
\newblock Probing quantum fluctuation theorems in engineered reservoirs.
\newblock {\em New Journal of Physics}, 19(10):103011, 2017.

\bibitem{monsel2018autonomous}
Juliette Monsel, Cyril Elouard, and Alexia Auffeves.
\newblock An autonomous quantum machine to measure the thermodynamic arrow of
  time.
\newblock {\em npj Quantum Information}, 4(1):1--9, 2018.

\bibitem{dressel2017arrow}
Justin Dressel, Areeya Chantasri, Andrew~N Jordan, and Alexander~N Korotkov.
\newblock Arrow of time for continuous quantum measurement.
\newblock {\em Physical review letters}, 119(22):220507, 2017.

\bibitem{manikandan2019time}
Sreenath~K Manikandan and Andrew~N Jordan.
\newblock Time reversal symmetry of generalized quantum measurements with past
  and future boundary conditions.
\newblock {\em Quantum studies: mathematics and foundations}, 6(2):241--268,
  2019.

\bibitem{manikandan2019fluctuation}
Sreenath~K Manikandan, Cyril Elouard, and Andrew~N Jordan.
\newblock Fluctuation theorems for continuous quantum measurements and absolute
  irreversibility.
\newblock {\em Physical Review A}, 99(2):022117, 2019.

\bibitem{risken1996fokker}
Hannes Risken.
\newblock Fokker-planck equation.
\newblock In {\em The Fokker-Planck Equation}, pages 63--95. Springer, 1996.

\bibitem{matos2020efficient}
Daniel Matos, Matthew~A Lane, Ian~J Ford, and Lev Kantorovich.
\newblock Efficient choice of colored noise in the stochastic dynamics of open
  quantum systems.
\newblock {\em Physical Review E}, 102(6):062134, 2020.

\bibitem{makri1995tensor}
Nancy Makri and Dmitrii~E Makarov.
\newblock Tensor propagator for iterative quantum time evolution of reduced
  density matrices. i. theory.
\newblock {\em The Journal of chemical physics}, 102(11):4600--4610, 1995.

\bibitem{yan2004hierarchical}
Yun-an Yan, Fan Yang, Yu~Liu, and Jiushu Shao.
\newblock Hierarchical approach based on stochastic decoupling to dissipative
  systems.
\newblock {\em Chemical physics letters}, 395(4-6):216--221, 2004.

\bibitem{suess2014hierarchy}
D~Suess, A~Eisfeld, and WT~Strunz.
\newblock Hierarchy of stochastic pure states for open quantum system dynamics.
\newblock {\em Physical review letters}, 113(15):150403, 2014.

\bibitem{orth2013nonperturbative}
Peter~P Orth, Adilet Imambekov, and Karyn Le~Hur.
\newblock Nonperturbative stochastic method for driven spin-boson model.
\newblock {\em Physical Review B}, 87(1):014305, 2013.

\bibitem{nakajima1958quantum}
Sadao Nakajima.
\newblock On quantum theory of transport phenomena: Steady diffusion.
\newblock {\em Progress of Theoretical Physics}, 20(6):948--959, 1958.

\bibitem{zwanzig1960ensemble}
Robert Zwanzig.
\newblock Ensemble method in the theory of irreversibility.
\newblock {\em The Journal of Chemical Physics}, 33(5):1338--1341, 1960.

\bibitem{mori1965transport}
Hazime Mori.
\newblock Transport, collective motion, and brownian motion.
\newblock {\em Progress of theoretical physics}, 33(3):423--455, 1965.

\bibitem{furutsu1964statistical}
Koichi Furutsu.
\newblock {\em On the statistical theory of electromagnetic waves in a
  fluctuating medium}.
\newblock United States Department of Commerce, 1964.

\bibitem{novikov1965functionals}
Evgenii~A Novikov.
\newblock Functionals and the random-force method in turbulence theory.
\newblock {\em Sov. Phys. JETP}, 20(5):1290--1294, 1965.

\bibitem{yan2016stochastic}
Yun-An Yan and Jiushu Shao.
\newblock Stochastic description of quantum brownian dynamics.
\newblock {\em Frontiers of Physics}, 11(4):1--24, 2016.

\bibitem{caldeira1983path}
Amir~O Caldeira and Anthony~J Leggett.
\newblock Path integral approach to quantum brownian motion.
\newblock {\em Physica A: Statistical mechanics and its Applications},
  121(3):587--616, 1983.

\bibitem{breuer1999stochastic}
Heinz-Peter Breuer, Bernd Kappler, and Francesco Petruccione.
\newblock Stochastic wave-function method for non-markovian quantum master
  equations.
\newblock {\em Physical Review A}, 59(2):1633, 1999.

\bibitem{breuer2004genuine}
Heinz-Peter Breuer.
\newblock Genuine quantum trajectories for non-markovian processes.
\newblock {\em Physical Review A}, 70(1):012106, 2004.

\bibitem{gisin1992quantum}
Nicolas Gisin and Ian~C Percival.
\newblock The quantum-state diffusion model applied to open systems.
\newblock {\em Journal of Physics A: Mathematical and General}, 25(21):5677,
  1992.

\bibitem{gisin1993quantum}
Nicolas Gisin and Ian~C Percival.
\newblock Quantum state diffusion, localization and quantum dispersion entropy.
\newblock {\em Journal of Physics A: Mathematical and General}, 26(9):2233,
  1993.

\bibitem{strunz1996linear}
Walter~T Strunz.
\newblock Linear quantum state diffusion for non-markovian open quantum
  systems.
\newblock {\em Physics Letters A}, 224(1-2):25--30, 1996.

\bibitem{brun2000continuous}
Todd~A Brun.
\newblock Continuous measurements, quantum trajectories, and decoherent
  histories.
\newblock {\em Physical Review A}, 61(4):042107, 2000.

\bibitem{percival1998quantum}
Ian Percival.
\newblock {\em Quantum state diffusion}.
\newblock Cambridge University Press, 1998.

\bibitem{clarke2021irreversibility}
Claudia~L. Clarke.
\newblock {\em Irreversibility Measures in a Quantum Setting}.
\newblock PhD thesis, UCL (University College London), 2021.

\bibitem{clarke2022entropy}
Claudia~L. Clarke and Ian~J. Ford.
\newblock In preparation.
\newblock 2022.

\bibitem{sachs1987physics}
Robert~G Sachs.
\newblock {\em The physics of time reversal}.
\newblock University of Chicago Press, 1987.

\end{thebibliography}

\end{document}